\documentclass[12pt,a4paper]{article}
\usepackage[british]{babel}
\usepackage{epsfig}
                                                                                
\begin{document}                                                                
\date{}

\title{
{\vspace{-3cm} \normalsize
\hfill \parbox{50mm}{DESY 01-214}}\\[25mm]
Supersymmetric Yang-Mills theory on the lattice
\footnote{Based on seminars given in 1999 to 2001.}\\[12mm]}
\author{I. Montvay                                           \\
Deutsches Elektronen-Synchrotron DESY,                       \\
Notkestr.\,85, D-22603 Hamburg, Germany}                                       
                                                                                
\newcommand{\be}{\begin{equation}}                                              
\newcommand{\ee}{\end{equation}}                                                
\newcommand{\half}{\frac{1}{2}}                                                 
\newcommand{\rar}{\rightarrow}                                                  
\newcommand{\lar}{\leftarrow}
\newcommand{\LCB}{\raisebox{-0.3ex}{\mbox{\LARGE$\left\{\right.$}}}
\newcommand{\RCB}{\raisebox{-0.3ex}{\mbox{\LARGE$\left.\right\}$}}}
                                                                                
\maketitle
\vspace*{1em}

\begin{abstract} \normalsize
 Recent development in numerical simulations of supersymmetric 
 Yang-Mills (SYM) theories on the lattice is reviewed.
\end{abstract}       

\newpage
\section{Introduction}\label{sec1}
 Supersymmetry seems to be a necessary ingredient of a quantum theory
 of gravity.
 Many of the possible extensions of the Standard Model beyond the
 presently known energy range are based on supersymmetry.
 It is generally assumed that the scale where supersymmetry becomes
 manifest is near to the presently explored electroweak scale and that
 the supersymmetry breaking is spontaneous.
 An attractive possibility for spontaneous supersymmetry breaking is to
 exploit non-perturbative mechanisms in supersymmetric gauge theories.
 This is the basis of a strong theoretical interest for investigating
 supersymmetry non-perturbatively.

 The motivation to investigate non-perturbative features of
 supersymmetric gauge theories is partly coming from the desire to
 understand relativistic quantum field theories better in general:
 the supersymmetric points in the parameter space of all quantum field
 theories are very special since they correspond to situations of a
 high degree of symmetry.
 The basic work of Seiberg and Witten \cite{SEIWIT} and other related
 papers showed that there is a possibility to approach non-perturbative
 questions in four dimensional quantum field theories by starting from
 exact solutions in some highly symmetric points and treat the symmetry
 breaking as a small perturbation.
 Beyond this, the knowledge of non-perturbative dynamics in
 supersymmetric quantum field theories can also be helpful in 
 understanding the greatest puzzle of the standard model, with or
 without supersymmetric extensions, namely the existence of a large
 number of seemingly free parameters.
 As we know from QCD, strong interactions in non-abelian gauge theories
 are capable to reproduce from a small number of input parameters a
 large number of dynamically generated parameters for quantities
 characterizing bound states.
 This is a possible solution also for the parameters of the standard
 model if new strong interactions are active beyond the electroweak
 symmetry breaking scale.

 The simplest supersymmetric gauge theory is the supersymmetric
 extension of Yang-Mills theory.
 It is the gauge theory of a massless Majorana fermion, called
 ``gaugino'', in the adjoint representation of the gauge group.
 The Euclidean action density of a gauge theory in the adjoint
 representation can be written as
\be  \label{eq101}
{\cal L}\; = \;\frac{1}{4}\, F^a_{\mu\nu}F^a_{\mu\nu}
+ \half\, \overline{\lambda}^a \gamma_\mu \left({\cal D}_\mu 
\lambda\right)^a + m_{\tilde{g}}\, \overline{\lambda}^a \lambda^a \ .
\ee
 Here $F^a_{\mu\nu}$ denotes the field strength tensor and $\lambda^a$
 is the Grassmannian fermion field, both with the adjoint representation
 index $a$.
 $m_{\tilde{g}}$ is the gaugino mass which has to be set equal to zero
 for supersymmetry.
 For a Majorana fermion $\lambda^a$ and $\overline{\lambda}^a$
 are not independent but satisfy
\be\label{eq102}
\overline{\lambda} = \lambda^T C \ ,
\ee
 with $C$ the charge conjugation Dirac matrix.
 This definition is based on the analytic continuation of Green's
 functions from Minkowski to Euclidean space \cite{MAJORANA}.

 The field theory defined by the Euclidean action (\ref{eq101}) has
 for $m_{\tilde{g}}=0$ a ``supersymmetry'' with respect to the
 infinitesimal transformations with a Grassmannian parameter
 $\epsilon$:
\be\label{eq103}
\delta A^a_\mu = 2i\overline{\epsilon}\gamma_\mu \lambda^a \ ,
\hspace*{3em}
\delta\lambda^a = -\sigma_{\mu\nu}F^a_{\mu\nu}\epsilon \ .
\ee
 It is easy to show that the change of the action density is a total
 derivative:
\be\label{eq104}
\delta{\cal L} = \partial_\mu \overline{j}_\mu\; \epsilon 
= \overline{\epsilon}\; \partial_\mu j_\mu
\ee
 where $j_\mu = -\half S_\mu $ and the ``supercurrent'' $S_\mu$ is
 defined as
\be\label{eq105}
S_\mu \equiv -F^a_{\rho\tau}\sigma_{\rho\tau}\gamma_\mu \lambda^a \ .
\ee
 Note that this is a Grassmannian Majorana current having both a 
 four-vector and a Dirac spinor index (this latter is not explicitly
 shown here).

 The existence of a supersymmetry in the above simple gauge field
 theory is at first sight surprising.
 It can be better understood in the general framework of supersymmetric
 field theories based on ``superfields'' \cite{SUSY}.
 In the Wess-Zumino gauge the action of Yang-Mills theory with $N=1$
 supersymmetry is conventionally given as
$$
\int d^4x\, d^2\theta\, {\rm Tr}(W^\alpha W_\alpha)
$$
\be\label{eq106}
= \int d^4x\, {\rm Tr}\left\{ -\half F_{\mu\nu}F^{\mu\nu}
+{i \over 2} F_{\mu\nu}\tilde{F}^{\mu\nu}
-i\lambda\sigma^\mu(D_\mu\bar{\lambda})
+i(D_\mu\bar{\lambda})\bar{\sigma}^\mu\lambda + D^2
\right\} \ ,
\ee
 where the first line is written in terms of the spinorial field
 strength superfield $W(x,\theta,\bar{\theta})_\alpha$ which depends on
 the four Minkowski-space coordinates $x_\mu,\; \mu=0,1,2,3$ and the
 anticommuting Weyl-spinor variables
 $\theta_\alpha,\bar{\theta}_{\dot\alpha}$ ($\alpha,\dot{\alpha}=1,2$).
 After performing the Grassmannian integration on $\theta$, one obtains
 the second form in terms of the component fields which are in this
 notation represented by Lie algebra elements.
 For instance, the field strength tensor $F_{\mu\nu}$ and its dual are
 defined as
\be\label{eq107}
F_{\mu\nu}(x) \equiv -ig F_{\mu\nu}^a(x) T_a \ , \hspace{3em}
\tilde{F}_{\mu\nu} \equiv \half\epsilon_{\mu\nu\rho\sigma} 
F^{\rho\sigma} \ .
\ee
 $\lambda,\bar{\lambda}$ in (\ref{eq106}) represent the Weyl components
 of the gaugino field.
 In the equation (\ref{eq101})-(\ref{eq103}) we denoted the
 corresponding Dirac spinors by the same letter $\lambda$.
 In the present paper we shall use, with very few exceptions, Dirac
 spinors therefore this will not give rise to confusion.

 The action in (\ref{eq106}) includes a $\Theta$-term, therefore it
 is natural to introduce the complex coupling
\be\label{eq108}
\tau \equiv \frac{\Theta}{2\pi} + \frac{4\pi i}{g^2}
\ee
 and then, with arbitrary $\Theta$, the $N=1$ SYM action becomes:
$$
\frac{1}{4\pi} \Im \left\{ \tau
\int d^4x\, d^2\theta\, {\rm Tr}(W^\alpha W_\alpha) \right\} 
$$
$$
= \frac{1}{g^2}\int d^4x\, {\rm Tr}\left[ -\half F_{\mu\nu}F^{\mu\nu}
-i\lambda\sigma^\mu(D_\mu\bar{\lambda})
+i(D_\mu\bar{\lambda})\bar{\sigma}^\mu\lambda + D^2
\right]
$$
\be\label{eq109}
+\frac{\Theta}{16\pi^2} \int d^4x\, 
{\rm Tr} \left[ F_{\mu\nu}\tilde{F}^{\mu\nu} \right] \ .
\ee
 Performing the trivial Gaussian integration over the auxiliary field
 $D$, going to Euclidean space and setting the $\Theta$-parameter to
 zero one obtains the action in the form (\ref{eq101}).

 The supersymmetry transformations in (\ref{eq103}) are relating the
 bosonic gauge field to the fermionic gaugino field.
 The above form of the transformation is ``on-shell'' because the
 auxiliary field $D$ is eliminated by the equation of motion $D=0$
 and it is ``non-linear'' due to the Wess-Zumino gauge fixing
 \cite{SUSY}.
 The realization of the supersymmetry in quantum field theory is
 not trivial because of the supersymmetry breaking introduced by
 the gauge fixing (for a recent discussion see \cite{RUSI}).
 Generally speaking one expects the existence of a renormalized
 supercurrent operator $S_{R\mu}$ satisfying the Ward-Takahashi-type
 identity
\be\label{eq110}
\partial_\mu S_{R\mu} = 2m_R \chi_R \ ,
\hspace{2em}
\left( \chi \equiv \half F^a_{\mu\nu}\sigma_{\mu\nu}\lambda^a \ ,
\hspace{2em} 
\chi_R = Z_\chi \chi \right) \ .
\ee
 Here $m_R$ is the renormalized gaugino mass and $Z_\chi$ is an
 appropriate multiplicative renormalization factor.
 The consequences of supersymmetry at $m_R=0$ can be obtained by
 considering different matrix elements of (\ref{eq110}).
 The non-vanishing right hand side of the above equation describes the
 ``soft breaking'' of supersymmetry due to non-zero gaugino mass.

\subsection{Open questions of SYM dynamics}\label{sec1.1}
 On the basis of its similarity to QCD one can assume that the basic
 features of SYM dynamics are similar to QCD: confinement of the coloured
 degrees of freedom and spontaneous chiral symmetry breaking.
 As in QCD, a central feature of low-energy dynamics is the realization
 of the global chiral symmetry.
 There is only a single Majorana adjoint ``flavour'' therefore the global
 chiral symmetry of $N=1$ SYM is $U(1)_\lambda$ which coincides with
 the so called {\em R-symmetry} generated by the transformations
\be\label{eq111}
\theta_\alpha^\prime = e^{i\varphi}\theta_\alpha \ , \hspace{2em}
\bar{\theta}_{\dot{\alpha}}^\prime = 
e^{-i\varphi}\bar{\theta}_{\dot{\alpha}} \ .
\ee
 This is equivalent to
\be\label{eq112}
\lambda_\alpha^\prime = e^{i\varphi}\lambda_\alpha \ , 
\hspace{1em}
\bar{\lambda}_{\dot{\alpha}}^\prime = 
e^{-i\varphi}\bar{\lambda}_{\dot{\alpha}} \ , 
\hspace{2em}
\lambda^\prime = e^{-i\varphi\gamma_5}\lambda \ ,
\hspace{1em}
\bar{\lambda}^\prime = \bar{\lambda}e^{-i\varphi\gamma_5}
\ee
 Here $\lambda_\alpha$, $\bar{\lambda}_{\dot{\alpha}}$ denote Weyl
 components and $\lambda$, $\bar{\lambda}$ without spinor indices are
 the Dirac spinor fields.

 The $U(1)_\lambda$-symmetry is anomalous.
 For definiteness, let us consider in what follows $SU(N_c)$ as gauge
 group when the corresponding axial current
 $J_\mu \equiv \bar{\lambda}\gamma_\mu\gamma_5\lambda$ satisfies
\be\label{eq113}
\partial^\mu J_\mu = \frac{N_c g^2}{32\pi^2} 
\epsilon^{\mu\nu\rho\sigma} F_{\mu\nu}^r F_{\rho\sigma}^r \ .
\ee
 Here $g$ denotes the gauge coupling.
 The anomaly leaves a $Z_{2N_c}$ subgroup of $U(1)_\lambda$ unbroken.
 This can be seen, for instance, by noting that the transformations
 in (\ref{eq112}) are equivalent to
\be\label{eq114}
m_{\tilde{g}} \to m_{\tilde{g}} e^{-2i\varphi\gamma_5} \ ,
\hspace{2em}
\Theta \to \Theta - 2N_c\varphi \ ,
\ee
 where $\Theta$ is the $\Theta$-parameter of gauge dynamics.
 Since $\Theta$ is periodic with period $2\pi$, for
 $m_{\tilde{g}}=0$ the $U(1)_\lambda$ symmetry is unbroken if
\be\label{eq115}
\varphi = \varphi_k \equiv \frac{k\pi}{N_c} \ , \hspace{2em}
(k=0,1,\ldots,2N_c-1) \ .
\ee
 For this statement it is essential that the topological charge is
 integer.

 The discrete global chiral symmetry $Z_{2N_c}$ is expected to be 
 spontaneously broken by the non-zero {\em gaugino condensate}
 $\langle \lambda\lambda \rangle \ne 0$ to $Z_2$ defined by 
 $\{\varphi_0,\varphi_{N_c}\}$ (note that $\lambda \to -\lambda$
 corresponding to $\varphi_{N_c}$ is a rotation).
 The consequence of this spontaneous chiral symmetry breaking pattern
 is the existence of a first order phase transition at zero gaugino
 mass $m_{\tilde{g}}=0$.
 For instance, in case of $N_c=2$ there exist two degenerate ground
 states with opposite signs of the gaugino condensate.
 The symmetry breaking is linear in  $m_{\tilde{g}}$, therefore the
 two ground states are exchanged at $m_{\tilde{g}}=0$ and there is a
 first order phase transition.

 For larger number of colours ($N_c > 2$) the phase structure is more
 involved.
 As an example, $N_c = 3$ gives rise to three degenerate ground states
 for the first order phase transition at the supersymmetry point.
 (For a first numerical study of SYM with $SU(3)$ see \cite{PISASU3}.)

 There are analytical predictions for the magnitude of the gaugino
 condensate based on instanton calculations.
 On general grounds the result has to be proportional to $\Lambda^3$
 where $\Lambda$ is the dynamical scale developed by {\em dimensional
 transmutation} which is defined by
\be\label{eq116}
\Lambda \equiv \mu e^{-{1}/{2\beta_0 g(\mu)^2}} \ ,\hspace{3em}
\beta_0 = \frac{3N_c}{16\pi^2} \ .
\ee
 Here $\mu$ is the scale belonging to the coupling $g(\mu)$.
 As usual, $\beta_0$ denotes the first coefficient of the
 $\beta$-function.
 In terms of $\Lambda^3$ the magnitude of the gaugino condensate can
 be written as
\be\label{eq117}
\langle \lambda^a_\alpha \lambda^{a\alpha} \rangle =  
C\Lambda^3 e^{2\pi ik/N_c} \ .
\ee
 The phase factor depending on the integer $k$ refers to the different
 ground states defined in (\ref{eq115}).
 The proportionality factor $C$ depends, of course, on the
 renormalization scheme where $\Lambda$ is defined.
 So called ``weak coupling'' instanton calculations
 \cite{AFDISE,NOSHVAZA,SHIVAI} imply that we have for $SU(N_c)$ gauge
 group $C=32\pi^2$ in the dimensional reduction scheme
 $\Lambda = \Lambda_{DR}$ \cite{FINPOU}.
 Another way of calculation using ``strong coupling'' gives, however,
 by a factor $2/((N_c-1)!(3N_c-1))^{1/N_c}$ different result
 \cite{NSVZ,ROSVEN,AMROVE,AKMRV}.
 (For a critique of the second method see \cite{HOKHLEMA}.)
 This discrepancy between the ``weak coupling'' and ``strong coupling''
 results could perhaps be due to the existence of an additional chiral
 symmetric vacuum state \cite{KOVSHI}.

 The predicted magnitude of the gaugino condensate (\ref{eq117}) can be
 checked in lattice Monte Carlo simulations.
 For this it is convenient to switch to the lattice $\Lambda$-parameter
 $\Lambda_{LAT}$.
 First one can use \cite{FINPOU}
\be\label{eq118}
\Lambda_{DR}/\Lambda_{\overline{MS}} = \exp\{-1/18\}
\ee
 and then for the lattice action which will be introduced in sections
 \ref{sec2} and \ref{sec3} one has \cite{WEISZ}
$$
\Lambda_{\overline{MS}}/\Lambda_{LAT} = \exp\left\{
-\frac{1}{\beta_0} \left[ \frac{1}{16N_c} - N_c P
+ \frac{N_c n_a}{2} P_3 \right]\right\} \ ,
$$
\be\label{eq119}
\beta_0 = \frac{N_c}{48\pi^2} (11-2n_a) \ ,\hspace{2em}
P = 0.0849780(1) \ ,\hspace{2em} P_3 = 0.0066960(1) \ .
\ee
 Here $n_a$ is the number of Majorana fermions in the adjoint
 representation, that is for SYM we have to set $n_a=1$.

 An important and interesting question is whether supersymmetry can be
 broken spontaneously or not.
 In pure SYM theory, without additional matter supermultiplets, this is
 not expected to occur.
 An argument for this is given by the non-zero value of the {\em Witten
 index} \cite{WITTEN}
\be\label{eq120}
w \equiv {\rm Tr\,}(-1)^F = n_{boson}-n_{fermion} \ ,
\ee
 which is equal to the difference of the number of bosonic minus
 fermionic states with zero energy.
 It is supposed not to change with the parameters of the theory.
 For SYM theory we have $w_{SYM}=N_c$, therefore there is no spontaneous
 supersymmetry breaking.
 (The $N_c$ ground states discussed above correspond to this Witten
 index.)

 In analogy with QCD, one expects that the spectrum of the SYM theory
 consists of colourless bound states formed out of the fundamental
 excitations, namely gluons and gluinos.
 (In this context we shall use the name ``gluino'' instead of the more
 general term ``gaugino''.)
 In the supersymmetric point at zero gluino mass these bound states
 should be organized in degenerate supersymmetry multiplets.
 For the description of lowest energy bound states one can try to use an
 effective field theory in terms of suitably chosen colourless composite
 operators.
 
 For $N=1$ SYM a low energy effective action was constructed by
 Veneziano and Yankielowicz (VY) \cite{VENYAN}.
 The composite operator appearing in the VY effective action is a
 chiral supermultiplet $S$ containing as component fields the
 expressions for the anomalies \cite{FERZUM}:
\be\label{eq121}
S \equiv A(y) + \sqrt{2} \theta\lambda(y) + \theta^2 F(y) \ ,
\ee
 where $y \equiv x+i\theta\sigma\bar{\theta}$.
 The scalar component of $S$ is proportional to the gluino bilinear
\be\label{eq122}
A \propto \lambda_\alpha \lambda^\alpha \ .
\ee
 The other components contain gluino-gluino and gluino-gluon
 combinations.
 Therefore, as far as a constituent picture is applicable to the bound
 states formed by strong interactions, the particle content of the
 lowest supersymmetry multiplet is: a pseudoscalar gluino-gluino bound
 state, a Majorana spinor gluon-gluino bound state and a scalar
 gluino-gluino bound state.
 In terms of $S$ the VY effective action has the form
\be\label{eq123}
S_{VY} = \frac{1}{\alpha}\int d^4x\, d^2\theta\, d^2\bar{\theta}\,
(S^\dagger S)^{1/3} +
\gamma\, [\int d^4x\, d^2\theta\, ( S \log \frac{S}{\Lambda} - S) 
+ {\rm h.c.}] \ .
\ee
 Here $\alpha$ and $\gamma$ are positive constants and $\Lambda$ is
 the mass parameter for the asymptotically free coupling which
 also appears in (\ref{eq117}).

 An interesting question is how the spectrum of glueballs, gluinoballs
 and gluino-glueballs is influenced by the soft supersymmetry breaking
 due to a non-zero gluino mass $m_{\tilde{g}} \ne 0$.
 For small $m_{\tilde{g}}$ it is possible to derive the coefficients
 of the terms linear in $m_{\tilde{g}}$ in the mass formulas
 \cite{EVHSSC}:
\begin{eqnarray}\label{eq124}
M_{\tilde{\eta}} &=& 
N_c\alpha\Lambda + \frac{40\pi^2|m_{\tilde{g}}|}{3N_c} \ ,
\nonumber \\
M_{\tilde{\chi}} &=& 
N_c\alpha\Lambda + \frac{48\pi^2|m_{\tilde{g}}|}{3N_c} \ ,
\nonumber \\
M_{\tilde{f}_0} &=& 
N_c\alpha\Lambda + \frac{56\pi^2|m_{\tilde{g}}|}{3N_c} \ .
\end{eqnarray}
 Here $\tilde{\eta}$, $\tilde{\chi}$ and $\tilde{f}_0$ denote the
 bound state with spin $0^-$, $\half$ and $0^+$, respectively.
 The range of applicability of the linear mass formulas in (\ref{eq124})
 is not known.

 The main assumption needed to derive the VY effective action
 (\ref{eq123}) is the choice of the chiral superfield $S$ as the
 dominant degree of freedom of low energy dynamics.
 Making a more general ansatz also containing gluon-gluon composites
 leads to a generalization and to two mixed supermultiplets in the
 low energy spectrum \cite{FAGASCH}.
 Other generalizations are conceivable.
 This shows that the description of low energy dynamics in N=1 SYM
 theory by chiral effective actions is less rigorous than, for instance,
 in QCD where we know from the Goldstone theorem that the light states
 are the members of the pseudoscalar meson multiplet.

 A general consequence of spontaneous chiral symmetry breaking is the
 existence of a first order phase transition at $m_{\tilde{g}}=0$.
 At this point the different ground states in (\ref{eq117}) are
 degenerate and a coexistence of the corresponding phases is possible.
 In a mixed phase situation, as usual at first order phase transitions,
 the different phases are separated by ``bubble wall'' interfaces.
 The {\em interface tension} of the walls can be exactly derived from
 the central extension of the $N=1$ SUSY algebra \cite{KOSHSM}.
 The result is that the energy density of the interface wall is related
 to the jump of the gluino condensate by
\be\label{eq125}
\epsilon = \frac{N_c}{8\pi^2} \left| \langle \lambda\lambda \rangle_1 
- \langle \lambda\lambda \rangle_2 \right| \ .
\ee
 Combining this with eq.~(\ref{eq117}) implies that the dimensionless
 ratio $\epsilon/|\langle \lambda\lambda \rangle|$ is predicted
 independently of the renormalization scheme.

 For transforming (\ref{eq117}) and (\ref{eq125}) to lattice units we
 need, in fact, the value of $a\Lambda_{LAT}$ at the particular values
 of interest of the lattice bare parameters ($a$ is, as usual, the
 lattice spacing).
 Before performing the lattice simulations this is, of course, not
 known.
 An order of magnitude estimate can be obtained from pure gauge theory
 (at infinitely heavy gluino) by noting that both for $N_c=2$ and
 $N_c=3$ we have for the lowest glueball mass $M$ \cite{GLUEBALL}
\be\label{eq126}
a\Lambda_{LAT} \simeq \frac{aM}{200} \ .
\ee
 Assuming this approximate relation also at the critical line for
 zero gluino mass, we can use eqs.~(\ref{eq117})-(\ref{eq126}) for
 estimating orders of magnitudes. 
 In the region where $aM={\cal O}(1)$ we obtain
\be\label{eq127}
a^3 |\langle \lambda\lambda \rangle_1 -
     \langle \lambda\lambda \rangle_2| = {\cal O}(1) \ ,
\hspace{3em}
a^3\epsilon = {\cal O}(10^{-1}) \ .
\ee
 As these numbers show, the predicted first order phase transition is,
 in fact, strong enough for a relatively easy observation in lattice
 simulations.

 A basic question about supersymmetry is the way it is realized in
 quantum field theory.
 The phenomenology of broken supersymmetry is based on the assumption
 that in the supersymmetric extension of the Standard Model
 supersymmetry is not anomalous: there is only ``soft breaking'' due to
 mass terms.
 Since SUSY is broken by the known regularizations (for a discussion on
 this see \cite{JACJON}), the question of a possible anomaly is
 legitimate.
 Non-perturbative effects may imply a {\em supersymmetry anomaly}
 \cite{CASHAM}.
 Investigating the behaviour of SUSY Ward-Takahashi identities in the
 continuum limit of lattice regularization can help to give an answer to
 this question.

\section{Lattice formulation}\label{sec2}
 In order to define the path integral for a Yang Mills theory with
 Majorana fermions in the adjoint representation, let us first
 consider the familiar case of Dirac fermions \cite{MM}.
 (Many of the notation conventions used in this paper are the same as
 in this reference.)
 Let us denote the Grassmanian fermion fields in the adjoint
 representation by $\psi^a_x$ and $\overline{\psi}^a_x$.
 Here Dirac spinor indices are omitted for simplicity and $a$ stands for
 the adjoint representation index ($a=1,..,N_c^2-1$ for SU($N_c$) ).
 The fermionic part of the Wilson lattice action is 
\be  \label{eq201}
S_f = \sum_x \LCB \overline{\psi}_x^a\psi_x^a
-K \sum_{\mu=1}^4 \left[
\overline{\psi}_{x+\hat{\mu}}^a V_{ab,x\mu}(1+\gamma_\mu)\psi_x^b
+\overline{\psi}_x^a V_{ab,x\mu}^T (1-\gamma_\mu)
\psi_{x+\hat{\mu}}^b \right] \RCB \ .
\ee
 Here $K$ is the hopping parameter, the Wilson parameter removing
 the fermion doublers in the continuum limit is fixed to $r=1$ and
 the matrix for the gauge-field link in the adjoint representation
 $V_{x\mu}$ is defined from the fundamental link variables $U_{x\mu}$
 according to
\be  \label{eq202}
V_{ab,x\mu} \equiv V_{ab,x\mu}[U] \equiv
2 {\rm Tr}(U_{x\mu}^\dagger T_a U_{x\mu} T_b)
= V_{ab,x\mu}^* =V_{ab,x\mu}^{-1T} \ .
\ee
 The generators $T_a \equiv \half \lambda_a$ satisfy the usual
 normalization ${\rm Tr\,}(\lambda_a\lambda_b)=\half$.
 In the simplest case of SU(2) ($N_c=2$) we have, of course,
 $T_a \equiv \half \tau_a$ with the isospin Pauli-matrices $\tau_a$.
 The normalization of the fermion fields in (\ref{eq201}) is the
 usual one for numerical simulations.
 The full lattice action is the sum of the pure gauge part and
 fermionic part: 
\be  \label{eq203}
S = S_g + S_f \ .
\ee
 The standard Wilson action for the SU($N_c$) gauge field $S_g$
 is a sum over the plaquettes
\be  \label{eq204}
S_g  =   \beta \sum_{pl}                                                  
\left( 1 - \frac{1}{N_c} {\rm Re\,Tr\,} U_{pl} \right) \ ,   
\ee
 with the bare gauge coupling given by $\beta \equiv 2N_c/g^2$.

 In order to obtain the lattice formulation of a theory with
 Majorana fermions let us note that out of a Dirac fermion field it
 is possible to construct two Majorana fields:
\be  \label{eq205}
\lambda^{(1)} \equiv \frac{1}{\sqrt{2}} ( \psi + C\overline{\psi}^T)
\ , \hspace{2em}
\lambda^{(2)} \equiv \frac{i}{\sqrt{2}} (-\psi + C\overline{\psi}^T)
\ee
 with the charge conjugation matrix $C$.
 These satisfy the Majorana condition
\be  \label{eq206}
\overline{\lambda}^{(j)} = \lambda^{(j)T} C
\hspace{3em} (j=1,2)  \ . 
\ee
 The inverse relation of (\ref{eq205}) is
\be  \label{eq207}
\psi = \frac{1}{\sqrt{2}} (\lambda^{(1)} + i\lambda^{(2)})
\ , \hspace{2em}
\psi_c \equiv C\overline{\psi}^T =
\frac{1}{\sqrt{2}} (\lambda^{(1)} - i\lambda^{(2)}) \ .
\ee
 In terms of the two Majorana fields the fermion action $S_f$ in
 eq.\ (\ref{eq201}) can be written as
\be  \label{eq208}
S_f = \half \sum_x \sum_{j=1}^2 \LCB
\overline{\lambda}_x^{(j)a}\lambda_x^{(j)a} 
-K \sum_{\mu=1}^4 \left[
\overline{\lambda}_{x+\hat{\mu}}^{(j)a} V_{ab,x\mu}
(1+\gamma_\mu)\lambda_x^{(j)b}
+\overline{\lambda}_x^{(j)a} V_{ab,x\mu}^T (1-\gamma_\mu)
\lambda_{x+\hat{\mu}}^{(j)b} \right] \RCB \ .
\ee

 For later purposes it is convenient to introduce the {\em fermion
 matrix}
\be  \label{eq209}
Q_{yd,xc} \equiv Q_{yd,xc}[U] \equiv
\delta_{yx}\delta_{dc} - K \sum_{\mu=1}^4 \left[
\delta_{y,x+\hat{\mu}}(1+\gamma_\mu) V_{dc,x\mu} +
\delta_{y+\hat{\mu},x}(1-\gamma_\mu) V^T_{dc,y\mu} \right] \ .
\ee
 Here, as usual, $\hat{\mu}$ denotes the unit vector in direction
 $\mu$.
 In terms of $Q$ we have
\be  \label{eq210}
S_f = \sum_{xc,yd} \overline{\psi}^d_y Q_{yd,xc} \psi^c_x
= \half\sum_{j=1}^2
\sum_{xc,yd} \overline{\lambda}^{(j)d}_y Q_{yd,xc} \lambda^{(j)c}_x \ ,
\ee
 and the fermionic path integral can be written as
\be  \label{eq211}
\int [d\overline{\psi} d\psi] e^{-S_f} = 
\int [d\overline{\psi} d\psi] e^{-\overline{\psi} Q \psi} = \det Q
= \prod_{j=1}^2 \int [d\lambda^{(j)}] 
e^{ -\half\overline{\lambda}^{(j)}Q\lambda^{(j)} } \ .
\ee
 This shows that the path integral over the Dirac fermion is the square
 of the path integral over the Majorana fermion and therefore
\be  \label{eq212}
\int [d\lambda] e^{ -\half\overline{\lambda} Q \lambda }
= \pm \sqrt{\det Q} \ .
\ee
 As one can see here, for Majorana fields the path integral involves
 only $[d\lambda^{(j)}]$ because of the Majorana condition in
 (\ref{eq206}).

 The relation (\ref{eq212}) leaves the sign on the right hand side
 undetermined.
 A unique definition of the path integral over a Majoran fermion field,
 including the sign, is given by
\be  \label{eq213}
\int [d\lambda] e^{ -\half\overline{\lambda} Q \lambda } 
= \int [d\lambda] e^{ -\half\lambda M \lambda } = {\rm Pf}(M)
\ee
 where $M$ is the antisymmetric matrix defined as
\be  \label{eq214}
M \equiv CQ = -M^T \ .
\ee

 The square root of the determinant in eq.\ (\ref{eq212}) is a
 {\em Pfaffian}.
 This can be defined for a general complex antisymmetric matrix
 $M_{\alpha\beta}=-M_{\beta\alpha}$ with an even number of dimensions
 ($1 \leq \alpha,\beta \leq 2N$) by a Grassmann integral as
\be  \label{eq215}
{\rm Pf}(M) \equiv
\int [d\phi] e^{-\half\phi_\alpha M_{\alpha\beta} \phi_\beta}
= \frac{1}{N! 2^N} \epsilon_{\alpha_1\beta_1 \ldots \alpha_N\beta_N}
M_{\alpha_1\beta_1} \ldots M_{\alpha_N\beta_N} \ .
\ee
 Here, of course, $[d\phi] \equiv d\phi_{2N} \ldots d\phi_1$, and 
 $\epsilon$ is the totally antisymmetric unit tensor.

 It is now clear that the fermion action for a Majorana fermion in the
 adjoint representation $\lambda^a_x$ can be defined by
\be  \label{eq216}
S_f \equiv \half \overline{\lambda} Q \lambda \equiv 
\half \sum_x \LCB \overline{\lambda}_x^a\lambda_x^a
-K \sum_{\mu=1}^4 \left[
\overline{\lambda}_{x+\hat{\mu}}^a V_{ab,x\mu}(1+\gamma_\mu)\lambda_x^b
+\overline{\lambda}_x^r V_{ab,x\mu}^T (1-\gamma_\mu)
\lambda_{x+\hat{\mu}}^b \right] \RCB \ .
\ee
 This together with (\ref{eq203})-(\ref{eq204}) gives a lattice action
 for the gauge theory of Majorana fermion in the adjoint representation.
 In order to achieve supersymmetry one has to tune the hopping parameter
 (bare mass parameter) $K$ to the {\em critical value} $K_{cr}(\beta)$
 in such a way that the mass of the fermion becomes zero.

 The path integral over $\lambda$ is defined by the Pfaffian
 ${\rm Pf}(CQ) = {\rm Pf}(M)$.
 By this definition the sign on the right hand side of 
 eq.\ (\ref{eq212}) is uniquely determined.
 The determinant $\det(Q)$ is real because the fermion matrix in
 (\ref{eq209}) satisfies
\be \label{eq217}
Q^\dagger = \gamma_5 Q \gamma_5 \ ,
\hspace{3em}
\tilde{Q} \equiv \gamma_5 Q = \tilde{Q}^\dagger \ .
\ee
 Moreover one can prove that $\det(Q)=\det(\tilde{Q})$ is always
 non-negative.
 This follows from the relations
\be\label{eq218}
CQC^{-1} = Q^T \ , \hspace{1em} B\tilde{Q}B^{-1} = \tilde{Q}^T \ ,
\ee
 with the charge conjugation matrix $C$ and $B \equiv C\gamma_5$.
 It follows that every eigenvalue of $Q$ and $\tilde{Q}$ is (at least)
 doubly degenerate.
 Therefore, with the real eigenvalues $\tilde{\lambda}_i$ of the
 Hermitean fermion matrix $\tilde{Q}$, we have
\be\label{eq219}
\det(Q) = \det(\tilde{Q}) = \prod_i \tilde{\lambda}_i^2 \geq 0 .
\ee
 Since according to the above discussion
\be\label{eq220}
\det(Q) = \det(M) = \left[ {\rm Pf}(M) \right]^2 \ ,
\ee
 the Pfaffian ${\rm Pf}(M)$ has to be real -- but it can have any sign.

 The relation between Majorana- and Dirac-fermions can also be
 exploited for the calculation of expectation values of Majorana
 fermion fields.
 For the Dirac fermion fields $\psi,\overline{\psi}$ we have, as is
 well known,
$$
\left\langle \psi_{y_1} \overline{\psi}_{x_1}
             \psi_{y_2} \overline{\psi}_{x_2}
\cdots       \psi_{y_n} \overline{\psi}_{x_n} \right\rangle =
Z^{-1} \int [d U] [d \overline{\psi}] [d \psi] \,
e^{-S_g[U]-\overline{\psi}Q\psi}
\psi_{y_1} \overline{\psi}_{x_1} \cdots 
\psi_{y_n} \overline{\psi}_{x_n}
$$
\be \label{eq221}
= Z^{-1} \int [d U]\, e^{-S_g[U]} \det Q[U] \sum_{z_1 \cdots z_n}
\epsilon^{z_1 z_2 \cdots z_n}_{y_1 y_2 \cdots y_n}
Q[U]^{-1}_{z_1x_1} Q[U]^{-1}_{z_2x_2} \cdots Q[U]^{-1}_{z_nx_n}
\ ,
\ee
with the antisymmetrizing unit tensor $\epsilon$ and
\be \label{eq222}
Z \equiv \int [d U] [d \overline{\psi}] [d \psi] \,
e^{-S_g[U]-\overline{\psi}Q\psi} = \int [d U]\, e^{-S_g[U]} \det Q[U]
\ .
\ee
 In order to express the expectation value of Majorana fermion fields
 by the matrix elements of the propagator $Q[U]^{-1}$, one can use
 again the doubling trick: one can identify, for instance,
 $\lambda \equiv \lambda^{(1)}$ and introduce another Majorana field
 $\lambda^{(2)}$, in order to obtain a Dirac field.
 Then from eqs.\ (\ref{eq205}) and (\ref{eq221}) we obtain
$$
\left\langle \lambda_{y_1} \overline{\lambda}_{x_1}
             \lambda_{y_2} \overline{\lambda}_{x_2}
\cdots       \lambda_{y_n} \overline{\lambda}_{x_n} \right\rangle =
Z_M^{-1} \int [d U]\, e^{-S_g[U]}\; {\rm Pf}(M[U]) (\det Q[U])^{-1}
$$
\be \label{eq223}
\cdot \int [d\overline{\psi}] [d\psi] \,
e^{-\overline{\psi} Q \psi} (\psi_{y_1}+C\overline{\psi}_{y_1})
(\overline{\psi}_{x_1}+\psi^T_{x_1}C) \cdots
(\psi_{y_n}+C\overline{\psi}_{y_n})
(\overline{\psi}_{x_n}+\psi^T_{x_n}C) 2^{-n} 
\ ,
\ee
 where now
\be \label{eq224}
Z_M \equiv \int [d U] \, e^{-S_g[U]}\; {\rm Pf}(M[U]) \ .
\ee

 The simplest example is $n=1$ when we have with $Q \equiv Q[U]$
$$
\left\langle \lambda_y \overline{\lambda}_x \right\rangle =
Z_M^{-1} \int [d U] e^{-S_g[U]}\; {\rm Pf}(M[U])\;
\half \left\{ Q^{-1}_{yx} + C^{-1}Q^{-1}_{xy}C \right\}
$$
\be \label{eq225}
= Z_M^{-1} \int [d U] e^{-S_g[U]}\; {\rm Pf}(M[U])\; Q^{-1}_{yx} \ .
\ee
 The important case $n=2$ can be expressed by six terms:
$$
\left\langle \lambda_{y_2} \overline{\lambda}_{x_2}
             \lambda_{y_1} \overline{\lambda}_{x_1} \right\rangle 
= Z_M^{-1} \int [d U] e^{-S_g[U]}\; {\rm Pf}(M[U])
$$
$$
\cdot \frac{1}{4} \sum_{z_1z_2} \left\{ \epsilon^{z_1z_2}_{y_1y_2}
Q^{-1}_{z_1x_1} Q^{-1}_{z_2x_2} +
C_{x_1}^{-1}\epsilon^{z_1z_2}_{x_1y_2} 
Q^{-1}_{z_1y_1} Q^{-1}_{z_2x_2}C_{y_1} + 
C_{x_2}^{-1}\epsilon^{z_1z_2}_{y_1x_2} 
Q^{-1}_{z_1x_1} Q^{-1}_{z_2y_2}C_{y_2} 
\right. 
$$
\be \label{eq226}
\left. + 
C_{x_1}^{-1}C_{x_2}^{-1} \epsilon^{z_1z_2}_{x_1x_2} 
Q^{-1}_{z_1y_1} Q^{-1}_{z_2y_2} C_{y_1}C_{y_2} -
C_{x_1}^{-1}\epsilon^{z_1z_2}_{y_1x_1} 
Q^{-1}_{z_1x_2} Q^{-1}_{z_2y_2}C_{y_2} - 
C_{x_2}^{-1}\epsilon^{z_1z_2}_{y_2x_2} 
Q^{-1}_{z_1x_1} Q^{-1}_{z_2y_1}C_{y_1} \right\} \ .
\ee
 The indices on the charge conjugation matrix $C$ show how the Dirac
 indices have to be contracted.

 Another way of expressing the expectation values of Majorana fermions
 is to consider the path integral with an external source $J_x$
 \cite{DGHV}:
\be \label{eq227}
Z_M[J] \equiv \int [d U] e^{-S_g[U]}\;
\int [d\lambda] e^{ -\half\lambda M \lambda - J\lambda} \ .
\ee
 After a shift of the Grassmanian integration variable
\be \label{eq228}
\lambda^\prime = \lambda + JM^{-1}
\ee
 one obtains
\be \label{eq229}
Z_M[J] = \int [d U] e^{-S_g[U]}\;
\int [d\lambda^\prime] e^{ -\half\lambda^\prime M \lambda^\prime
 - \half JM^{-1}J}
=  \int [d U] e^{-S_g[U]}\; {\rm Pf}(M)\; e^{- \half JM^{-1}J} \ .
\ee
 Expectation values of the form (\ref{eq225}), (\ref{eq226}) can now
 be obtained by differentiation with respect to the source $J$.

\subsection{Improved actions}\label{sec2.1}
 The lattice action defined by (\ref{eq203}), (\ref{eq204}) and
 (\ref{eq216}) is not unique.
 One expects that there is a wide class of actions belonging to the same
 {\em universality class} which all reproduce the same theory in the
 continuum limit.
 This ambiguity can be used for {\em improvement} of some desired
 property, for instance, faster approach to the continuum limit and/or
 less symmetry breaking.

 Direct improvement of the supersymmetry of the lattice action seems to
 be rather hard in gauge theories because the gauge field and the
 fermions are introduced on the lattice very differently.
 Nevertheless, in Wess-Zumino type scalar-fermion models it is possible
 to achieve {\em perfect supersymmetry} on the lattice with respect to
 supersymmetry type transformations \cite{BIETEN}.
 (See also \cite{CATKAR}.)

 Another possibility of improvement is to apply a lattice formulation
 for fermions where the zero gaugino mass is achieved without fine
 tuning.
 The preferred way to do this is based on {\em domain wall fermions}
 \cite{KAPSCH} where an auxiliary fifth dimension is introduced
 which contains a four dimensional domain wall supporting a massless
 fermion.
 A similar alternative is based on {\em overlap fermions}
 \cite{NEUBERGER,HOIZNI}.
 In these formulations an additional advantage is the improved chiral
 symmetry.
 In fact, these lattice fermions are prominent representatives of the
 Ginsparg-Wilson realization of chiral symmetry on the lattice
 \cite{GINWIL}.

 An important advantage of Ginsparg-Wilson fermions is due to their
 clean spectral properties: for zero mass the eigenvalues are in the
 complex plane on a circle crossing the real axis at zero.
 For non-zero mass the crossing is shifted by the mass in lattice units.
 The product of eigenvalues (determinant or Pfaffian) cannot be
 negative if the mass is non-negative.
 This is in contrast with Wilson-type fermions where the fluctuation
 of eigenvalues implies the possibility of negative eigenvalues
 also for (small) positive masses.
 This is the origin of a possible ``sign problem'' with Wilson-type
 lattice fermions (see section \ref{sec3.2}).

 Most of the numerical simulations up to now used the simple Wilson-type
 lattice action in (\ref{eq216}).
 These will be discussed in detail in sections~\ref{sec4}-\ref{sec6}.
 Recently there has been, however, also a Monte Carlo study of SYM
 theory based on the domain wall fermion action \cite{FKV}.
 In this work an evidence was found for the formation of a gaugino
 condensate in the chiral limit at zero gaugino mass. 

\section{Algorithm for numerical simulations}\label{sec3}
 In order to perform Monte Carlo simulations of SYM theory one needs
 a positive measure on the gauge field which allows for importance
 sampling of the path integral.
 Therefore the sign of the Pfaffian can only be taken into account by
 reweighting.
 According to (\ref{eq220}) the absolute value of the Pfaffian is the
 non-negative square root of the determinant therefore the effective
 gauge field action is \cite{CURVEN}:
\be \label{eq301}
S_{CV} = \beta\sum_{pl} \left( 1-\half{\rm Tr\,}U_{pl} \right)
- \half\log\det Q[U] \ .
\ee
 The factor $\half$ in front of $\log\det Q$ shows that we effectively
 have a flavour number $N_f=\half$ of adjoint fermions.
 The omitted sign of the Pfaffian can be taken into account by reweighting
 the expectation values according to
\be\label{eq302}
\langle {\cal O} \rangle = \frac{\langle {\cal O}\; 
{\rm sign Pf}(M)\rangle_{CV}} {\langle {\rm sign Pf}(M)\rangle_{CV}} \ ,
\ee
 where $\langle \ldots \rangle_{CV}$ denotes expectation values with
 respect to the effective gauge action $S_{CV}$.
 This may give rise to a {\em sign problem} which will be discussed in
 section \ref{sec3.2}.

 The fractional power of the determinant corresponding to (\ref{eq301})
 can be reproduced, for instance, by the hybrid molecular dynamics
 algorithm \cite{HMD} which is, however, a finite step size algorithm
 where the step size has to be extrapolated to zero.
 This adds another necessary extrapolation to three: to infinite volume,
 to small gaugino mass and to the continuum limit.
 An ``exact'' algorithm where the step size extrapolation is absent
 is the {\em two-step multi-boson} (TSMB) algorithm \cite{TSMB}.

\subsection{Two-step multi-boson algorithm}\label{sec3.1}
 The multi-boson algorithms for dynamical (``unquenched'') fermion
 simulations \cite{MB} are based on polynomial approximations as
\be\label{eq303}
\left|\det(Q)\right|^{N_f} =
\left\{\det(Q^\dagger Q) \right\}^{N_f/2}
\simeq \frac{1}{\det P_n(Q^\dagger Q)} \ ,
\ee
 where $N_f$ is the number of flavours (in our case $N_f=\half$) and the
 polynomial $P_n$ satisfies
\be\label{eq304}
\lim_{n \to \infty} P_n(x) = x^{-N_f/2}
\ee
 in an interval $[\epsilon,\lambda]$ covering the spectrum of
 $Q^\dagger Q = \tilde{Q}^2$.
 (Here $\tilde{Q}$ is the hermitean fermion matrix defined in
 (\ref{eq217}).)

 For the multi-boson representation of the determinant one uses
 the roots of the polynomial $r_j,\; (j=1,\ldots,n)$
\be\label{eq305}
P_n(Q^\dagger Q) = P_n(\tilde{Q}^2) =
r_0 \prod_{j=1}^n (\tilde{Q}^2 - r_j) \ .
\ee
 Assuming that the roots occur in complex conjugate pairs, one can
 introduce the equivalent forms
\be\label{eq306}
P_n(\tilde{Q}^2)
= r_0 \prod_{j=1}^n [(\tilde{Q} \pm \mu_j)^2 + \nu_j^2]
= r_0 \prod_{j=1}^n (\tilde{Q}-\rho_j^*) (\tilde{Q}-\rho_j)
\ee
 where $r_j \equiv (\mu_j+i\nu_j)^2$ and $\rho_j \equiv \mu_j + i\nu_j$.
 With the help of complex boson (pseudofermion) fields $\Phi_{jx}$
 one can write
\be\label{eq307}
\prod_{j=1}^n\det[(\tilde{Q}-\rho_j^*) (\tilde{Q}-\rho_j)]^{-1} \propto
\int [d\Phi]\; \exp\left\{ -\sum_{j=1}^n \sum_{xy}
\Phi_{jy}^+\, [(\tilde{Q}-\rho_j^*) (\tilde{Q}-\rho_j)]_{yx}\,
\Phi_{jx} \right\} \ .
\ee
 Since for a finite polynomial of order $n$ the approximation in
 (\ref{eq304}) is not exact, one has to extrapolate the results to
 $n\to\infty$.

 The difficulty for small fermion masses in large physical volumes is
 that the {\em condition number} $\lambda/\epsilon$ becomes very large
 ($10^4-10^6$) and very high orders $n = {\cal O}(10^3)$ are needed for
 a good approximation.
 This requires large storage and the autocorrelation of the gauge
 configurations becomes very bad since the autocorrelation length is a
 fast increasing function of $n$.
 One can achieve substantial improvements on both these problems by
 introducing a two-step polynomial approximation:
\be\label{eq308}
\lim_{n_2 \to \infty} P^{(1)}_{n_1}(x)P^{(2)}_{n_2}(x) =
x^{-N_f/2} \ , \hspace{3em}
x \in [\epsilon,\lambda] \ .
\ee
 The multi-boson representation is only used for the first
 polynomial $P^{(1)}_{n_1}$ which provides a first crude approximation
 and hence the order $n_1$ can remain relatively low.
 The correction factor $P^{(2)}_{n_2}$ is realized in a stochastic
 {\em noisy correction step} with a global accept-reject condition
 during the updating process.
 In order to obtain an exact algorithm one has to consider in this case
 the limit $n_2\to\infty$.

 In the two-step approximation scheme for $N_f$ flavours of fermions
 the absolute value of the determinant is represented as
\be\label{eq309}
\left|\det(Q)\right|^{N_f} \;\simeq\;
\frac{1}{\det P^{(1)}_{n_1}(\tilde{Q}^2)
\det P^{(2)}_{n_2}(\tilde{Q}^2)} \ .
\ee
 The multi-boson updating with $n_1$ scalar pseudofermion fields
 is performed by heatbath and overrelaxation sweeps for the boson
 fields and Metropolis (or heatbath and overrelaxation) sweeps over
 the gauge field.
 After an update sweep over the gauge field a global accept-reject
 step is introduced in order to reach the distribution of gauge field
 variables $[U]$ corresponding to the right hand side of
 (\ref{eq309}).
 The idea of the noisy correction is to generate a random vector
 $\eta$ according to the normalized Gaussian distribution
\be \label{eq310}
\frac{e^{-\eta^\dagger P^{(2)}_{n_2}(\tilde{Q}[U]^2)\eta}}
{\int [d\eta] e^{-\eta^\dagger P^{(2)}_{n_2}(\tilde{Q}[U]^2)\eta}}  \ ,
\ee
 and to accept the change $[U] \to [U^\prime]$ with probability
\be \label{eq311}
\min\left\{ 1,A(\eta;[U^\prime] \lar [U]) \right\} \ ,
\ee
 where
\be \label{eq312}
A(\eta;[U^\prime] \lar [U]) =
\exp\left\{-\eta^\dagger P^{(2)}_{n_2}(\tilde{Q}[U^\prime]^2)\eta 
           +\eta^\dagger P^{(2)}_{n_2}(\tilde{Q}[U]^2)\eta\right\}\ .
\ee

 The Gaussian noise vector $\eta$ can be obtained from $\eta^\prime$
 distributed according to the simple Gaussian distribution
\be \label{eq313}
\frac{e^{-\eta^{\prime\dagger}\eta^\prime}}
{\int [d\eta^\prime] e^{-\eta^{\prime\dagger}\eta^\prime}}
\ee
 by setting it equal to
\be \label{eq314}
\eta = P^{(2)}_{n_2}(\tilde{Q}[U]^2)^{-\half} \eta^\prime  \ .
\ee

 In order to obtain the inverse square root on the right hand side of
 (\ref{eq314}), one can proceed with a polynomial approximation
\be \label{eq315}
 P^{(3)}_{n_3}(x) \simeq P^{(2)}_{n_2}(x)^{-\half} \ , \hspace{1em}
x \in [0,\lambda] \ .
\ee
 This is a relatively easy approximation because
 $P^{(2)}_{n_2}(x)^{-\half}$ is not singular at $x=0$, in contrast to
 the function $x^{-N_f/2}$.
 A practical way to obtain $P^{(3)}$ is to use some approximation scheme
 for the inverse square root.
 The best possibility is to use a Newton iteration
\be \label{eq316}
P^{(3)}_{k+1} = \half \left( P^{(3)}_k + \frac{1}{P^{(3)}_k P^{(2)}} 
\right) \ , \hspace{3em} k=0,1,2,\ldots  \ . 
\ee
 The second term on the right hand side can be evaluated by a polynomial
 approximation as for $P^{(2)}$ in (\ref{eq308}) with $N_f=0$ and
 $P^{(1)} \to P^{(3)}_k P^{(2)}$.
 The iteration in (\ref{eq316}) is fast converging and allows for an
 iterative procedure stopped by a prescribed precision.
 A starting polynomial $P^{(3)}_0$ can be obtained, for instance,
 from (\ref{eq308}) with $N_f \to -\half N_f$ or by the formula
 \cite{OVERLAP}
\be \label{eq317}
P^{(3)} \simeq \frac{1}{K} \sum_{s=1}^K \frac{1}
{P^{(2)} \cos^2\frac{\pi}{2K}(s-\half) + \sin^2\frac{\pi}{2K}(s-\half)} \ .
\ee

 The TSMB algorithm becomes exact only in the limit of infinitely high
 polynomial order: $n_2\to\infty$ in (\ref{eq308}).
 Instead of investigating the dependence of expectation values on $n_2$
 by performing several simulations, it is better to fix some high order
 $n_2$ for the simulation and perform another correction in the
 ``measurement'' of expectation values by still finer polynomials.
 This is done by {\em reweighting} the configurations.

 Before discussing this {\em measurement correction} for general
 number of flavours $N_f$ let us note that $N_f=2$, and any other even
 number, is a special case.
 This is because, due to $P^{(2)}(x) \simeq [x\,P^{(1)}(x)]^{-1}$, the
 noisy correction can be performed in this case by an iterative
 inversion, as first introduced in \cite{BORFOR}.
 Consequently, one can proceed without a measurement correction.
 In another multi-boson scheme \cite{FREJAN} (polynomial hybrid Monte
 Carlo) a measurement correction is needed but for even $N_f$ it can be
 performed by an iterative inversion.

 The measurement correction for general $N_f$ is based on a polynomial
 approximation $P^{(4)}_{n_4}$ which satisfies
\be\label{eq318}
\lim_{n_4 \to \infty} P^{(1)}_{n_1}(x)P^{(2)}_{n_2}(x)P^{(4)}_{n_4}(x) =
x^{-N_f/2} \ , \hspace{3em}
x \in [\epsilon^\prime,\lambda] \ .
\ee
 The interval $[\epsilon^\prime,\lambda]$ can be chosen by convenience,
 for instance, such that $\epsilon^\prime=0,\lambda=\lambda_{max}$,
 where $\lambda_{max}$ is an absolute upper bound of the eigenvalues of
 $Q^\dagger Q=\tilde{Q}^2$.
 (In practice, instead of $\epsilon^\prime=0$, it is more effective to
 take $\epsilon^\prime > 0$ and determine the eigenvalues below
 $\epsilon^\prime$ and the corresponding correction factors explicitly
 \cite{SPECTRUM}.)
 In this case the limit $n_4\to\infty$ is exact on an arbitrary gauge
 configuration.
 For the evaluation of $P^{(4)}_{n_4}$ one can use $n_4$-independent
 recursive relations \cite{POLYNOM}, which can be stopped by observing
 the required precision of the result.
 After reweighting the expectation value of a quantity ${\cal O}$ is
 given by
\be\label{eq319}
\langle {\cal O} \rangle = \frac{
\langle {\cal O} \exp{\{\eta^\dagger[1-P^{(4)}_{n_4}(Q^\dagger Q)]\eta\}}
\rangle_{U,\eta}}
{\langle  \exp{\{\eta^\dagger[1-P^{(4)}_{n_4}(Q^\dagger Q)]\eta\}}
\rangle_{U,\eta}} \ ,
\ee
 where $\eta$ is a simple Gaussian noise like $\eta^\prime$ in
 (\ref{eq313}).
 Here $\langle\ldots\rangle_{U,\eta}$ denotes an expectation value
 on the gauge field sequence, which is obtained in the two-step process
 described in the previous subsection, and on a sequence of independent
 $\eta$'s.
 The expectation value with respect to the $\eta$-sequence can be
 considered as a Monte Carlo updating process with the trivial action
 $S_\eta \equiv \eta^\dagger\eta$.
 The length of the $\eta$-sequence on a fixed gauge configuration can,
 in principle, be arbitrarily chosen.
 In praxis it has to be optimized for obtaining the smallest possible
 errors.
 If the second polynomial gives a good approximation the correction
 factors do not practically change the expectation values.
 In this reweighting step the sign of the Pfaffian can also be included
 according to (\ref{eq302}) and then one has
\be\label{eq320}
\langle {\cal O} \rangle = \frac{
\langle {\cal O}\; {\rm sign Pf}(M) \;
\exp{\{\eta^\dagger[1-P^{(4)}_{n_4}(Q^\dagger Q)]\eta\}} 
\rangle_{U,\eta}}
{\langle {\rm sign Pf}(M) \;
\exp{\{\eta^\dagger[1-P^{(4)}_{n_4}(Q^\dagger Q)]\eta\}} 
\rangle_{U,\eta}} \ .
\ee

 A useful generalization of the two-step multi-boson algorithm is to
 allow for the fermion action in the multi-boson update step to differ
 from the true fermion action one wants to simulate \cite{VARYACT}.
 For instance, for improved actions one can take a simplified version
 in the multi-boson update step.
 Random variation of the parameters in the update step may also be used
 to decrease autocorrelations.

\subsection{The ``sign problem''}\label{sec3.2}

 The Pfaffian resulting from the Grassmannian path integrals for
 Majorana fermions (\ref{eq213}) is an object similar to a determinant
 but less often used.
 As shown by (\ref{eq215}), ${\rm Pf}(M)$ is a polynomial of the
 matrix elements of the $2N$-dimensional antisymmetric matrix $M=-M^T$.
 Basic relations are \cite{PFAFFIAN}
\be\label{eq321}
M = P^T J P ,\hspace{3em}  {\rm Pf}(M)=\det(P) \ ,
\ee
 where $J$ is a block-diagonal matrix containing on the diagonal
 $2\otimes2$ blocks equal to $\epsilon=i\sigma_2$ and otherwise zeros.
 Let us note that from these relations the second equality in
 eq.~(\ref{eq220}) immediately follows.

 The form of $M$ required in (\ref{eq321}) can be achieved by a
 procedure analogous to the Gram-Schmidt orthogonalization and, by
 construction, $P$ is a triangular matrix (see \cite{SPECTRUM}).
 This gives a numerical procedure for the computation of $P$ and the
 determinant of $P$ gives, according to (\ref{eq321}), the Pfaffian
 ${\rm Pf}(M)$.
 Since $P$ is triangular, the calculation of $\det(P)$ is, of course,
 trivial.

\begin{figure}[th]
\vspace*{0.2cm}
\begin{center}
\epsfig{file=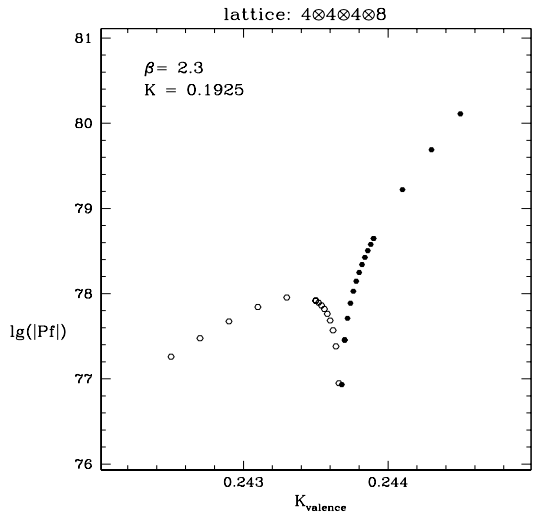,
        width=7.5cm,height=7.0cm,
        angle=0}
\vspace*{-0.5cm}
\parbox{12cm}{\caption{\label{fig01}\em
 The absolute value of the pfaffian on a $4^3\cdot8$ configuration as a
 function of the hopping parameter in $M$ (and $\tilde{Q}$).
 Open points stand for ${\rm Pf}(M)>0$, full ones for ${\rm Pf}(M)<0$.}}
\end{center}
\vspace*{0.2cm}
\end{figure}

 This procedure can be used for a numerical determination of the
 Pfaffian on small lattices \cite{BOULDER}.
 For an example see figure \ref{fig01} which also shows a sign change of
 ${\rm Pf}(M)$ as a function of the hopping parameter $K$.

 On lattices larger than, say, $4^3 \cdot 8$ the computation becomes
 cumbersome due to the large storage requirements.
 This is because one has to store a full $\Omega \otimes \Omega$ matrix,
 with $\Omega$ being the number of lattice points multiplied by the
 number of spinor-colour indices (equal to $4(N_c^2-1)$ for the adjoint
 representation of ${\rm SU}(N_c)$).
 The difficulty of computation is similar to a computation of the
 determinant of $Q$ with $LU$-decomposition.

\begin{figure}[htb]
\vspace*{0.2cm}
\begin{center}
\epsfig{file=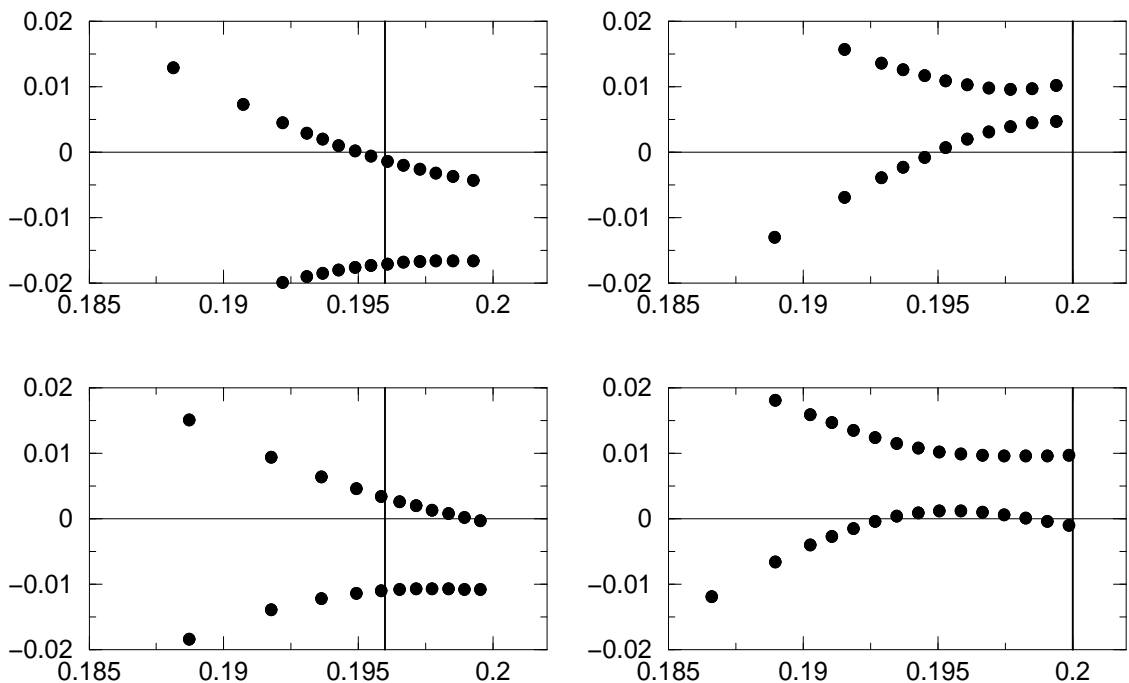,
        width=15cm,height=9cm}
\end{center}
\vspace*{-0.5cm}
\begin{center}
\parbox{15cm}{\caption{\label{fig02}\em
 The spectral flow of the hermitean fermion matrix $\tilde{Q}$ for some
 specific configurations on $6^3 \cdot 12$ lattice at $\beta=2.3$.
 The value of $K$ in the simulation is displayed by a vertical line.}}
\end{center}
\vspace*{0.2cm}
\end{figure}

 Fortunately, in order to obtain the sign of the Pfaffian occurring in
 the measurement reweighting formulas (\ref{eq302}), (\ref{eq320}) one
 can proceed without a full calculation of the value of the Pfaffian.
 The method is to monitor the sign changes of ${\rm Pf}(M)$ as a
 function of the hopping parameter $K$.
 According to (\ref{eq219}), the hermitean fermion matrix for the gaugino
 $\tilde{Q}$ has doubly degenerate real eigenvalues therefore
\be\label{eq322}
\det M = \det \tilde{Q} = \prod_{i=1}^{\Omega/2} \tilde{\lambda}_i^2 \ ,
\ee
 where $\tilde{\lambda}_i$ denotes the eigenvalues of $\tilde{Q}$.
 This implies
\be\label{eq323}
|{\rm Pf}(M)| =  \prod_{i=1}^{\Omega/2} |\tilde{\lambda}_i| \ ,
\hspace{2em} \Longrightarrow \hspace{2em}
{\rm Pf}(M) =  \prod_{i=1}^{\Omega/2} \tilde{\lambda}_i \ .
\ee
 The first equality trivially follows from (\ref{eq220}).
 The second one is the consequence of the fact that ${\rm Pf}(M)$ is
 a polynomial in $K$ which cannot have discontinuities in any of its
 derivatives.
 Therefore if, as a function of $K$, an eigenvalue $\tilde{\lambda}_i$
 (or any odd number of them) changes sign the sign of ${\rm Pf}(M)$
 has to change, too.
 Since at $K=0$ we have ${\rm Pf}(M)=1$, the number of sign changes
 between $K=0$ and the actual value of $K$, where the dynamical fermion
 simulation is performed, determines the sign of ${\rm Pf}(M)$.
 This means that one has to determined the flow of the eigenvalues of
 $\tilde{Q}$ through zero \cite{SPECTRALFLOW}.
 Examples of the spectral flow taken from data of the Monte Carlo
 simulations of the DESY-M\"unster Collaboration \cite{SPECTRUM} are
 shown in figure \ref{fig02}.

 The spectral flow method is well suited for the calculation of the
 sign of the Pfaffian in SYM theory.
 An important question is the frequency and the effects of
 configurations with negative sign.
 A strongly fluctuating Pfaffian sign is a potential danger for the
 effectiveness of the Monte Carlo simulation because cancellations
 can occur resulting in an unacceptable increase of statistical errors.
 The experience of the DESY-M\"unster Collaboration shows, however,
 that below the critical line $K_{cr}(\beta)$ corresponding to zero
 gaugino mass ($m_{\tilde{g}}=0$) negative Pfaffians practically never
 appear \cite{SPECTRUM,WTIB,WTI}.
 Above the critical line several configurations with negative Pfaffian
 have been observed but their r\^ole has not yet been cleared up to
 now.
 Since supersymmetry is expected to be realized in the continuum limit
 at  $m_{\tilde{g}}=0$, the negative signs of the Pfaffian can be
 avoided if one takes the zero gaugino mass limit from
 $m_{\tilde{g}} > 0$ corresponding to $K < K_{cr}$.
 In this sense there is no ``sign problem'' in SYM with Wilson fermions
 which would prevent a Monte Carlo investigation.

\begin{figure}[htb]
\vspace*{0.2cm}
\begin{center}
\epsfig{file=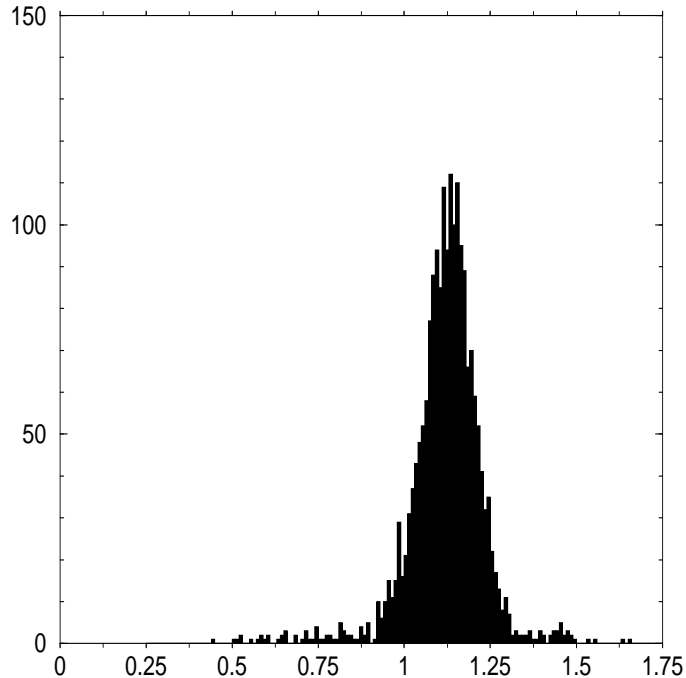,
        width=9cm,height=9cm,angle=0}
\end{center}
\vspace*{-0.5cm}
\begin{center}
\parbox{12cm}{\caption{\label{fig03}\em
 The distribution of reweighting factors at $\beta=2.3,\; K=0.1940$ on 
 $12^3\cdot 24$ lattice.}}
\end{center}
\vspace*{0.2cm}
\end{figure}

\begin{figure}[htb]
\vspace*{0.2cm}
\begin{center}
\epsfig{file=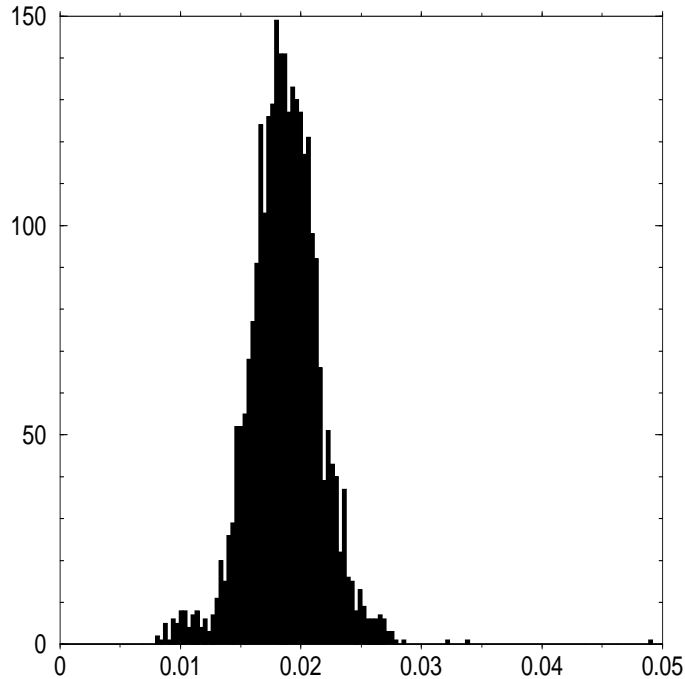,
        width=9cm,height=9cm,angle=0}
\end{center}
\vspace*{-0.5cm}
\begin{center}
\parbox{12cm}{\caption{\label{fig04}\em
 The distribution of smallest eigenvalues of $\tilde{Q}^2$ at
 $\beta=2.3,\; K=0.1940$ on  $12^3\cdot 24$ lattice.}}
\end{center}
\vspace*{0.2cm}
\end{figure}

 The presence or absence of negative Pfaffians in a sample of gauge
 configurations produced in Monte Carlo simulations can be easily
 seen even without the application of the spectral flow method.
 In case of sign changes the distribution of the reweighting factors
 in (\ref{eq319}) shows a pronounced tail reaching down to zero
 \cite{DESYSWANSEA}.
 If this tail is absent, as in the example of a run \cite{WTI} shown by
 figure \ref{fig03}, then there are no negative Pfaffians.
 Alternatively, one has to observe the distribution of smallest
 eigenvalues of $\tilde{Q}^2$ which also shows a tail down to zero
 if there are negative Pfaffians in the Monte Carlo sample.
 The absence of a tail shows that there are no negative Pfaffians.
 This is illustrated by figure \ref{fig04} belonging to the same
 configuration sample as figure \ref{fig03}.

 Concerning this ``sign problem'' let us note that a very similar
 phenomenon exists also in QCD because the Wilson-Dirac determinant of
 a single quark flavour can also have a negative sign.
 Under certain circumstances the sign of the quark determinant plays
 an important r\^ole.
 This is the case, for instance, at large quark chemical potential in a
 QCD-like model with SU(2) colour and staggered quarks in the adjoint
 representation which has recently been studied by the DESY-Swansea
 Collaboration \cite{DESYSWANSEA}.
 This investigation also revealed an interesting feature of the TSMB
 algorithm, namely its ability to easily change the sign of eigenvalues
 of the hermitean fermion matrix (and hence the sign of the determinant
 or Pfaffian).
 This is in contrast to algorithms based on finite difference molecular
 dynamics equations as, for instance, the hybrid Monte Carlo \cite{HMC}
 or HMD \cite{HMD} algorithms.

\section{Discrete chiral symmetry breaking}\label{sec4}
 It has been discussed in section \ref{sec1.1} that in SYM theory with
 gauge group $SU(N_c)$ the discrete global chiral symmetry $Z_{2N_c}$ is
 expected to be spontaneously broken by a non-zero gaugino condensate
 $\langle \lambda\lambda \rangle \ne 0$.
 The consequence of this chiral symmetry breaking pattern is the
 existence of a first order phase transition at zero gaugino
 mass $m_{\tilde{g}}=0$.

\begin{figure}[htb]
\vspace*{0.7cm}
\begin{center}
\epsfig{file=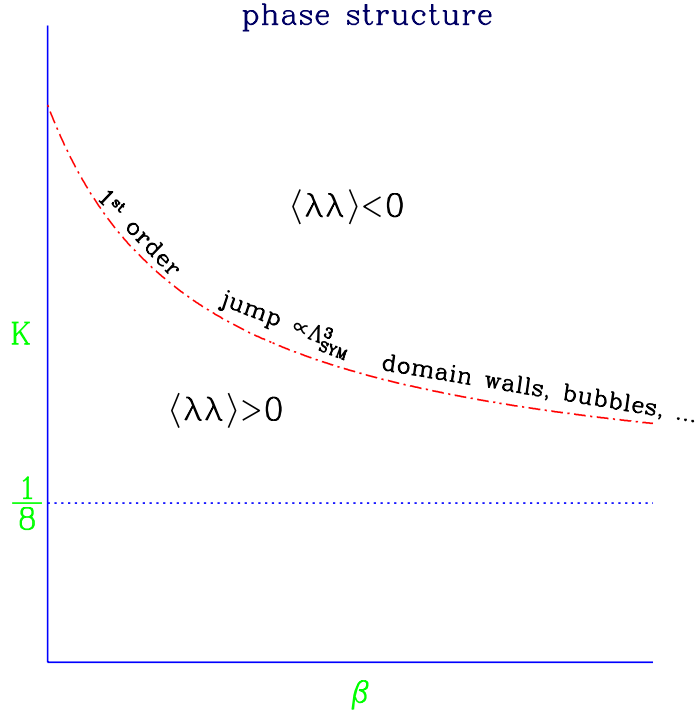,
        width=10cm,height=10cm,angle=0,
        bbllx=80pt,bblly=190pt,bburx=680pt,bbury=760pt}
\end{center}
\vspace*{-1.7cm}
\begin{center}
\parbox{12cm}{\caption{\label{fig05}\em
 Expected phase structure of Yang-Mills theory with a Majorana
 fermion in adjoint representation in the ($\beta,K$)-plane.
 The dashed-dotted line $K=K_{cr}(\beta)$ is a first order phase
 transition (or cross-over) at zero gaugino mass.}}
\end{center}
\vspace*{-0.2cm}
\end{figure}

 In case of the lattice formulation based on the lattice action in
 (\ref{eq203})-(\ref{eq204}), (\ref{eq216}) the bare parameters are
 the gauge coupling $\beta \equiv 2N_c/g^2$ and the hopping parameter
 $K$.
 In the plane ($\beta,K$) there is a line corresponding to zero
 gaugino mass $K_{cr}(\beta)$ where a first order phase transition
 occurs.
 Therefore the expected phase structure is the one shown in
 figure~\ref{fig05}.
 In fact, the theoretical expectation of a genuine phase transition
 refers to the continuum limit $\beta \to \infty$.
 It is possible that for finite $\beta$ there is a cross-over which
 becomes a real first order phase transition only at $\beta=\infty$.

 The DESY-M\"unster Collaboration performed a first lattice
 investigation of gaugino condensation in SYM theory with SU(2) gauge
 group \cite{DISCHIRAL}.
 The distribution of $\langle\lambda\lambda\rangle$ has been studied
 for fixed gauge coupling $\beta=2.3$ as a function of the hopping
 parameter $K$, which determines the bare gaugino mass, on
 $6^3 \cdot 12$ lattice.

 A first order phase transition shows up on small to moderately
 large lattices as metastability expressed by a two-peak structure
 in the distribution of the gaugino condensate.
 By tuning $K$ one should be able to achieve that the two peaks are
 equal (in height or area).
 This is a possible definition of the phase transition point in finite
 volumes.
 By increasing the volume the tunneling between the two ground states
 becomes less and less probable and at some point practically
 impossible.
\begin{figure*}
\begin{flushleft}
\hspace*{0.08\textwidth}
\includegraphics[width=0.36\textwidth]
{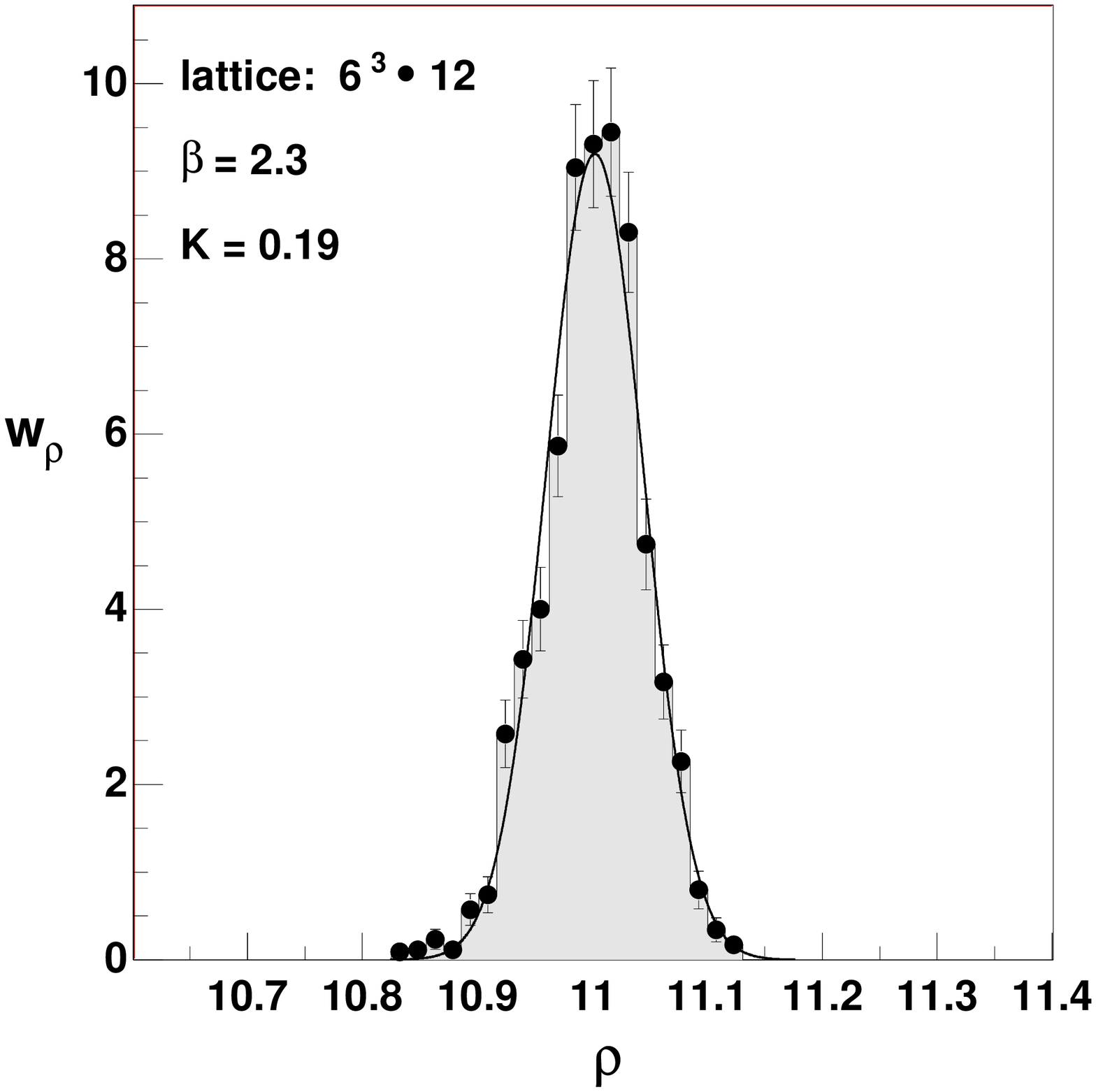}
\end{flushleft}
\vspace*{-0.287\textheight}
\begin{flushright}
\includegraphics[width=0.36\textwidth]
{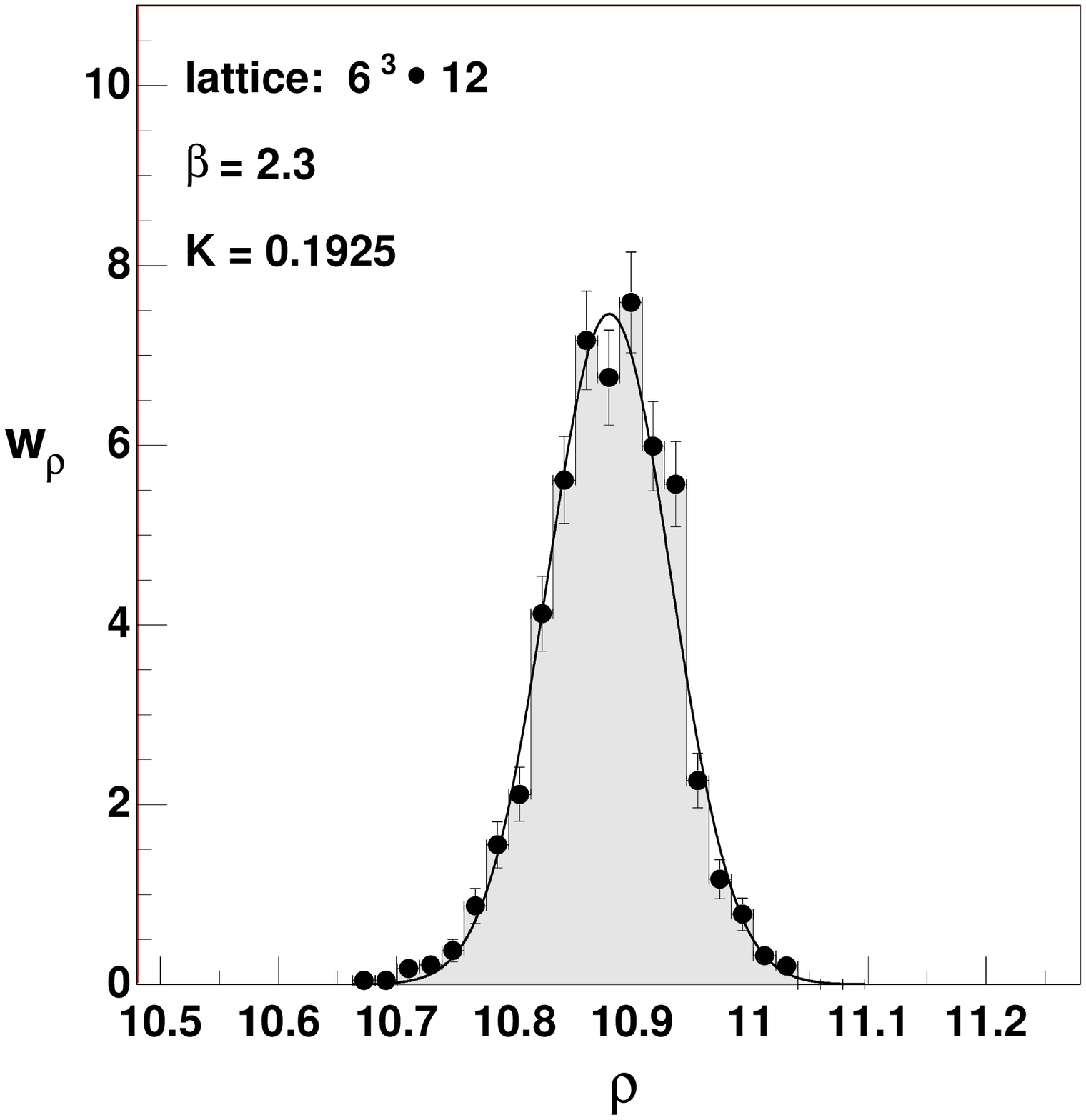}
\hspace*{0.10\textwidth}
\end{flushright}
\begin{flushleft}
\hspace*{0.08\textwidth}
\includegraphics[width=0.36\textwidth]
{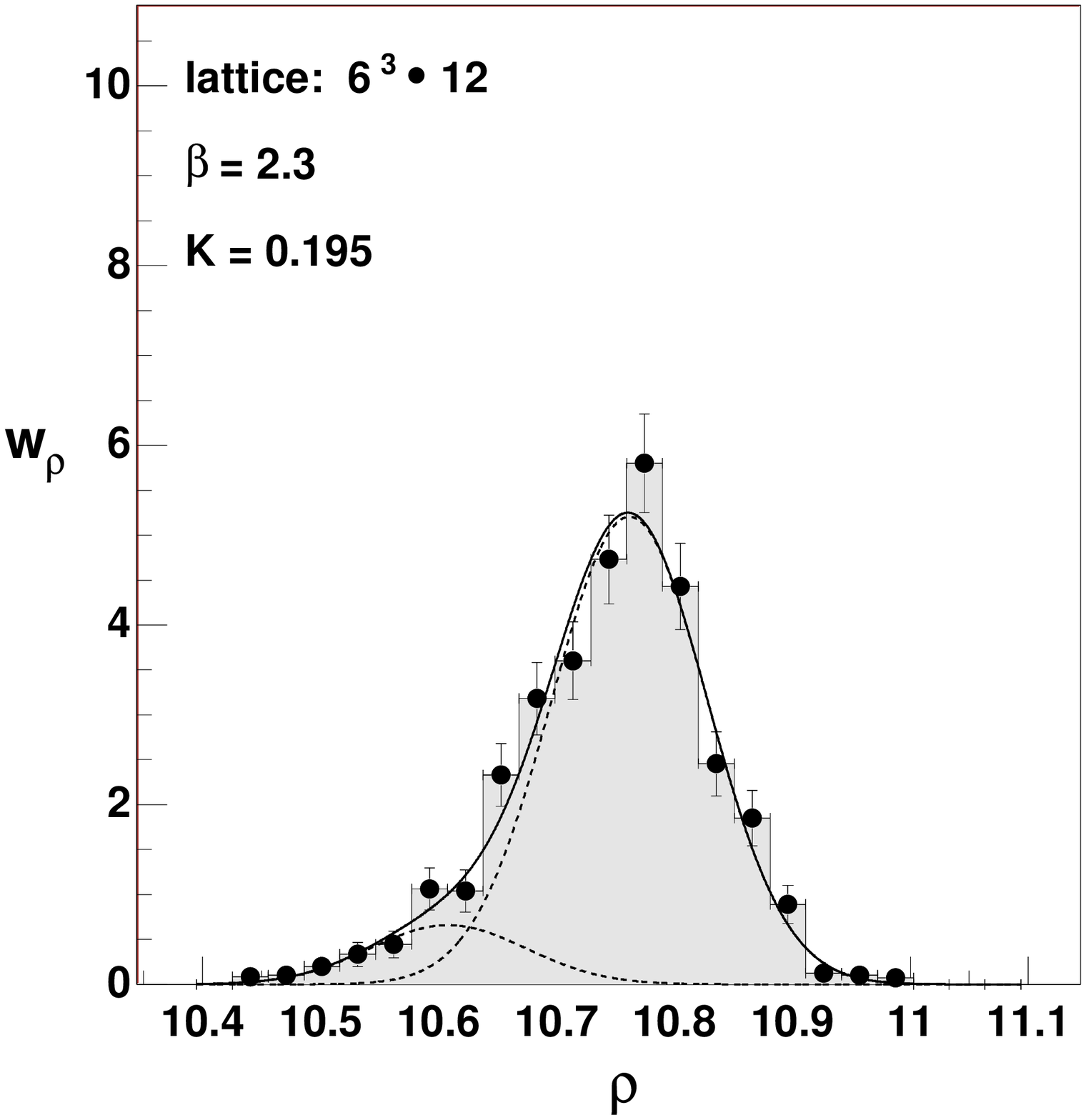}
\end{flushleft}
\vspace*{-0.287\textheight}
\begin{flushright}
\includegraphics[width=0.36\textwidth]
{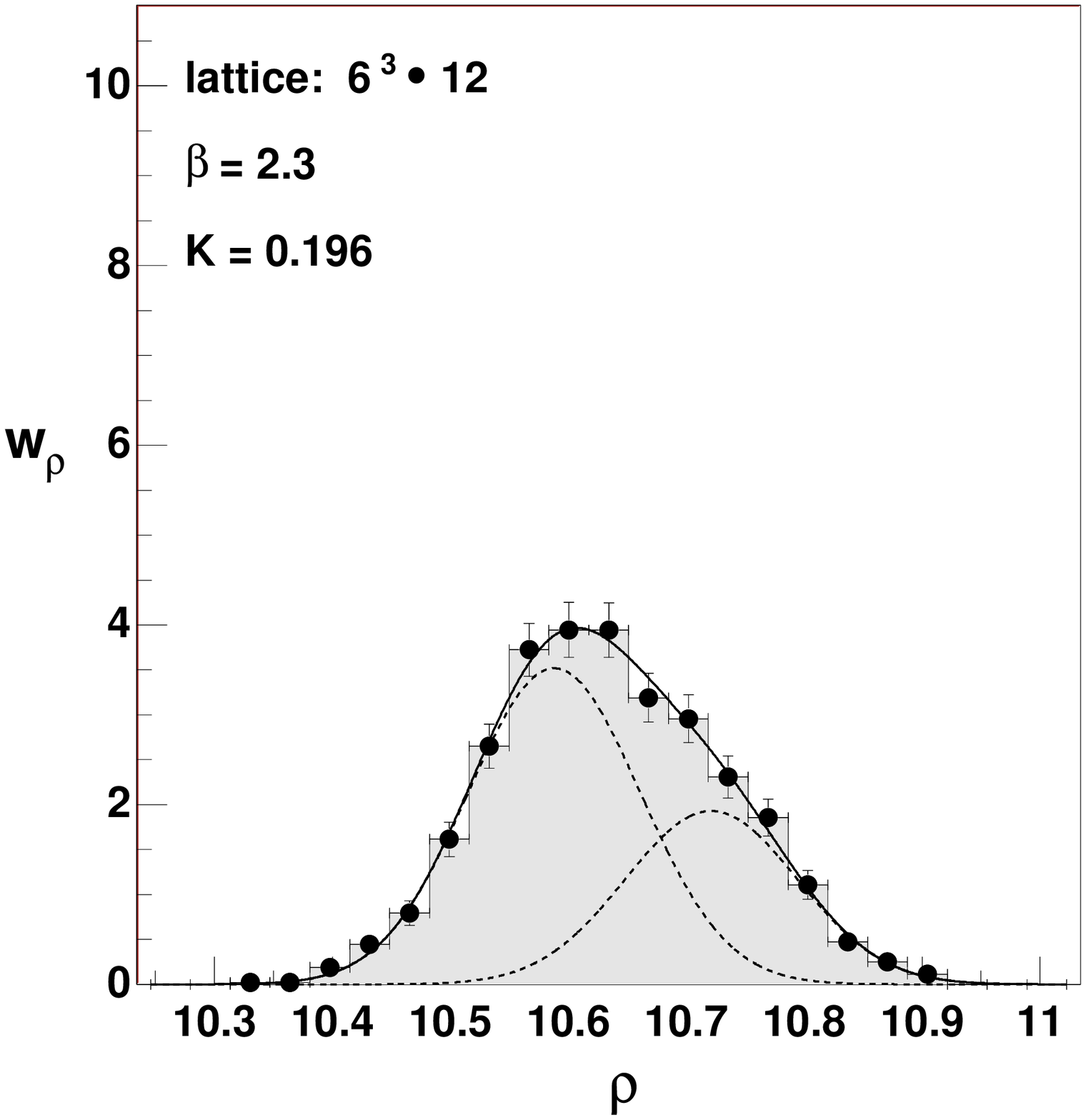}
\hspace*{0.10\textwidth}
\end{flushright}
\begin{flushleft}
\hspace*{0.08\textwidth}
\includegraphics[width=0.36\textwidth]
{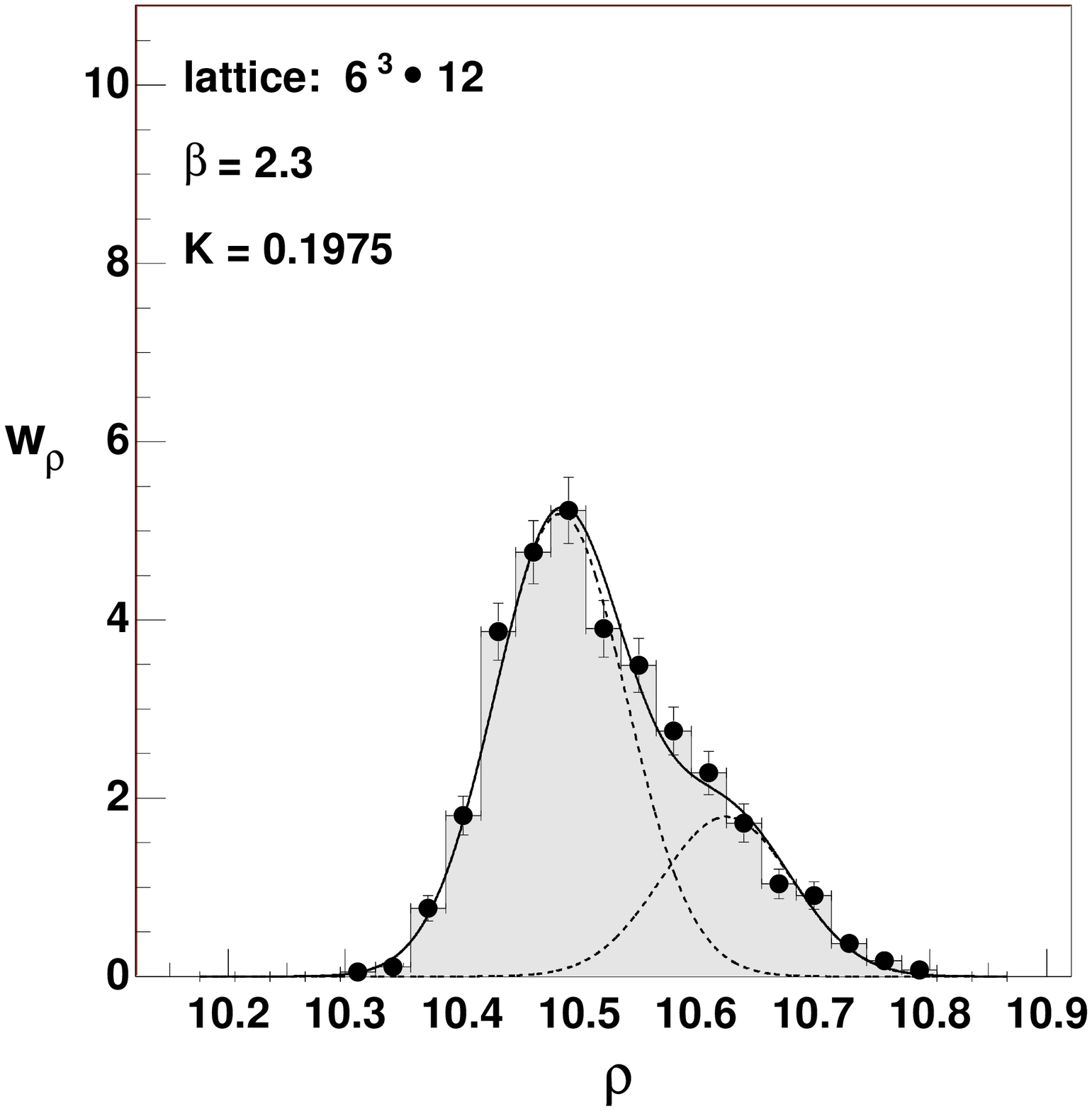}
\end{flushleft}
\vspace*{-0.287\textheight}
\begin{flushright}
\includegraphics[width=0.36\textwidth]
{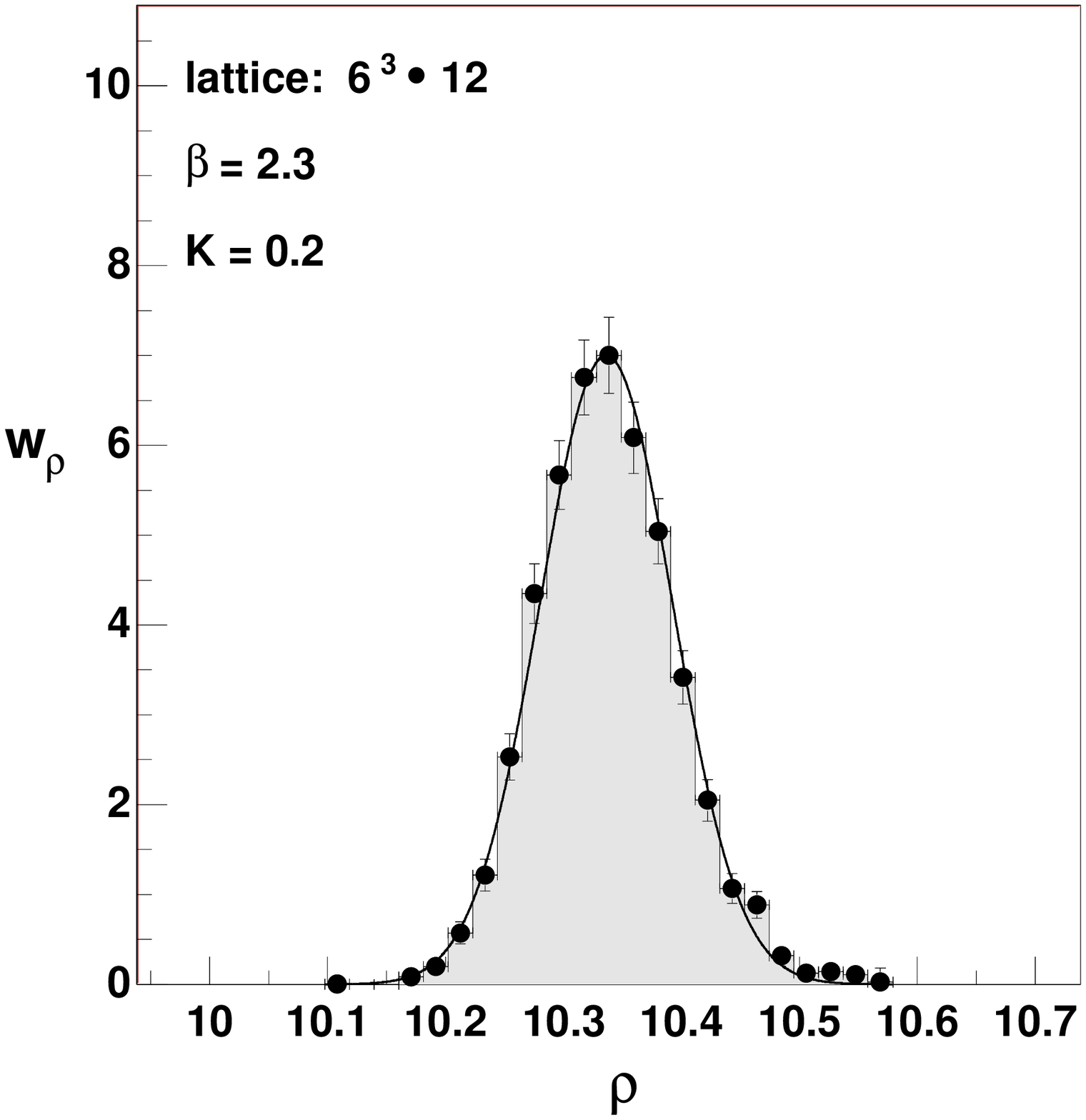}
\hspace*{0.10\textwidth}
\end{flushright}
\vspace*{-0.7cm}
\begin{center}
\parbox{14cm}{\caption{\label{fig06}\em
 The probability distributions of the gaugino condensate
 $\rho\equiv\langle\lambda\lambda\rangle$ for different
 hopping parameters at $\beta=2.3$ on $6^3 \cdot 12$ lattice.
 The dased lines show the Gaussian components.}}
\end{center}
\end{figure*}

 The observed distributions are shown in figure~\ref{fig06}.
 One can see that the distributions cannot always be described by a
 single Gaussian which would correspond to a single phase.
 Parameters of possible fits are collected in table~\ref{tab01}.
\begin{table}[ht]
\begin{center}
\parbox{15cm}{\caption{\label{tab01}\em
 Fit parameters of the distributions of gaugino condensate corresponding
 to figure~\protect\ref{fig06}.
 The weight of the first peak is $p_1$, the mean values are $\mu_{1,2}$
 and the common width of the Gaussians is $\sigma$.
 The statistical errors in last digits are given in parentheses.
}}
\end{center}
\begin{center}
\begin{tabular}{| l | l | l | l | l | l |}
\hline
\multicolumn{1}{|c|}{$K$}              &
\multicolumn{1}{|c|}{$p_1$}            &  
\multicolumn{1}{|c|}{$\mu_1$}          &  
\multicolumn{1}{|c|}{$\mu_2$}          & 
\multicolumn{1}{|c|}{$\sigma$}         &
\multicolumn{1}{|c|}{$\chi^2/d.o.f.$}  \\
\hline\hline
0.19   &   1.0   &  11.0023(26) & \multicolumn{1}{|c|}{-} & 
0.0423(16) & 27.9/20 \\
\hline
0.1925 &   1.0   &  10.8807(30) & \multicolumn{1}{|c|}{-} & 
0.0524(17) & 25.9/20 \\
\hline
0.195  & 0.89(7) &  10.762(30)  &  10.608(30)  &  0.066(7)  & 16.5/18 \\
\hline
0.196  & 0.35(7) &  10.722(11)  &  10.588(11)  &  0.073(3)  &  5.7/18 \\
\hline
0.1975 & 0.26(5) &  10.626(17)  &  10.484(17)  &  0.056(4)  & 19.5/18 \\
\hline
0.2    &   0.0   & \multicolumn{1}{|c|}{-} &
10.3363(37) & 0.0562(18) & 21.4/20 \\
\hline\hline
\end{tabular}
\end{center}
\end{table}
 As the table shows, in the region $0.195 \leq K \leq 0.1975$ only 
 two-Gaussian fits work.
 Outside this region single Gaussian fits describe the data well.
 For increasing $K$ (decreasing bare gaugino mass) the weights shift
 from the Gaussian at larger $\rho\equiv\langle\lambda\lambda\rangle$ to
 the one with smaller $\rho$, as expected.
 The two Gaussians represent the contributions of the two phases on this
 lattice.
 The position of the phase transition (or cross-over) on the
 $6^3 \cdot 12$ lattice is at $K_0=0.1955 \pm 0.0005$.
 According to table~\ref{tab01} the jump of the order parameter is
 $\Delta\rho \equiv \mu_1 - \mu_2 \simeq 0.15$.

 The two-phase structure can also be searched for in pure gauge field
 variables as the plaquette or longer Wilson loops.
 It turns out that the distributions of Wilson loops are rather
 insensitive.
 They can be well described by single Gaussians with almost constant
 variance in the whole range $0.19 \leq K \leq 0.2$ \cite{DISCHIRAL}.

 As shown by figure~\ref{fig06} and table~\ref{tab01}, on this small
 lattice the two peaks corresponding to two phases are not well
 separated.
 It is possible that at this $\beta$ value there is no genuine first
 order phase transition at all -- only a cross-over.
 The question of first order phase transition versus cross-over can be
 decided by studying the volume dependence on larger lattices.
 After extracting the jump of the gaugino condensate in the large
 (infinite) volume limit one has to study its behaviour in the
 continuum limit $\beta\to\infty$ along the line $K_{cr}(\beta)$ in
 figure~\ref{fig05}.
 The renormalized gaugino condensate can be obtained in the continuum
 limit by additive and multiplicative renormalizations:
\be\label{eq401}
\langle \bar{\lambda}_x \lambda_x \rangle_{R(\mu)} = Z(a\mu)
\left[ \langle \bar{\lambda}_x \lambda_x \rangle - b_0(a\mu) \right] \ .
\ee
 The presence of the additive shift in the gaugino condensate
 $b_0(a\mu)$ implies that the value of its jump at $m_{\tilde{g}}=0$
 is easier available than the value itself.

 The renormalization factor $Z$ in (\ref{eq401}) is expected to be of
 order ${\cal O}(1)$ at bare parameter values which can be reached by
 numerical simulations.
 For obtaining it one can use non-perturbative renormalization schemes
 (for a recent review see \cite{SINT}).
 Performing perturbative calculations in lattice regularization can
 also give useful information \cite{TANIG,SUSYWTP}.

\section{Confinement and particle spectrum}\label{sec5}

\subsection{Fundamental static potential}\label{sec5.1}
 The potential between static colour sources in gauge field theory is a
 physically interesting quantity because it is characteristic for the
 dynamics of the gauge field.
 The external static soursec can be put in any representation of the
 gauge group.
 If the sources are in the fundamental representation we have to do with
 the {\em fundamental static potential} between {\em static quarks}.

 For a model containing dynamical matter fields in the fundamental
 representation, as is the case for QCD, the charge of static quarks
 will be screened.
 The potential then approaches a constant at large distances.
 The string tension $\sigma$, which is the asymptotic slope of the
 potential for large distances, vanishes accordingly.
 On the other hand, if only matter fields in the adjoint representation
 of the gauge group are present, as in the case of N=1 SYM theory,
 confinement of static quarks and a positive $\sigma$ are expected.

 The DESY-M\"unster Collaboration has determined the static quark
 potential and the string tension for N=1 SUSY Yang-Mills theory with
 gauge group SU(2) in Monte Carlo simulations \cite{SPECTRUM}.
 The starting point of numerical work are expectation values of
 rectangular Wilson loops $\langle W(R,T) \rangle$ of spatial length
 $R$ and time length $T$.
 From the Wilson loops the potential can be found via
\be\label{eq501}
V(R) = \lim_{T \to \infty} V(R,T) \ ,
\ee
 where
\be\label{eq502}
V(R,T) = 
\log \langle W(R,T) \rangle - \log \langle W(R,T+1) \rangle \ .
\ee
 The potential $V(R)$ is obtained through a fit of the form
\be\label{eq503}
V(R,T) = V(R) + c_1(R) e^{-c_2(R) T}.
\ee
 The string tension $\sigma$ is finally obtained by fitting the
 potential according to
\be\label{eq504}
V(R) = V_0 - \frac{\alpha}{R} + \sigma R.
\ee

\begin{figure}[htb]
\vspace*{0.2cm}
\begin{center}
\epsfig{file=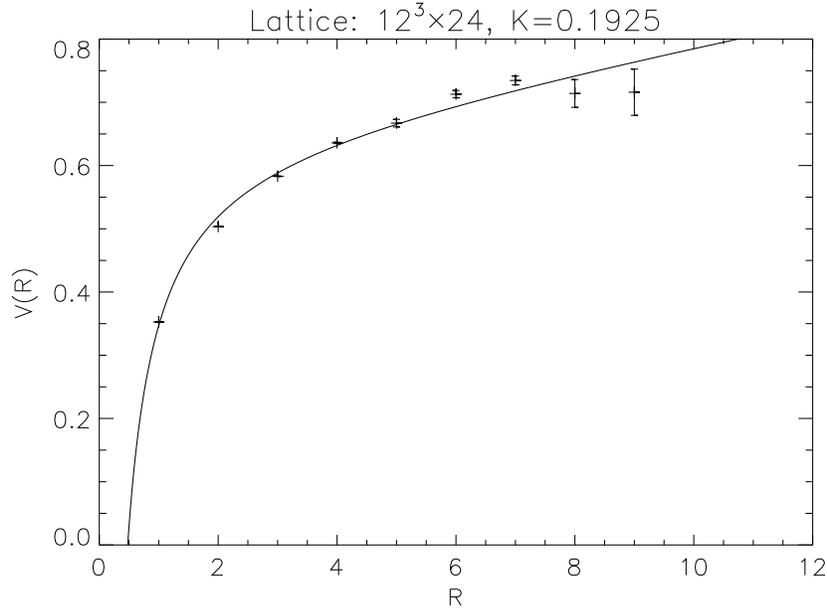,
        width=12cm}
\end{center}
\vspace*{-1.2cm}
\begin{center}
\parbox{12cm}{\caption{\label{fig07}\em
 The static quark potential $V(R)$ on a $12^3 \cdot 24$ lattice at
 $\beta=2.3$, $K=0.1925$. The line is a fit with a Coulomb plus a linear term,
 fitted over the range $1 \leq R \leq 6$.}}
\end{center}
\vspace*{-0.5cm}
\end{figure}

 An example of the static quark potential on $12^3 \cdot 24$ lattice
 at $\beta=2.3$, $K=0.1925$ is shown in figure~\ref{fig07}.
 A collection of the results for the string tension on lattices with
 spatial extension $L$ is:
\begin{eqnarray}\label{eq505}
a \sqrt{\sigma} &=& 0.22(1) \hspace{5mm}  \mbox{for}\ K=0.1900, \ L=8,
\nonumber \\
a \sqrt{\sigma} &=& 0.21(1) \hspace{5mm}  \mbox{for}\ K=0.1925, \ L=8,
\nonumber \\
a \sqrt{\sigma} &=& 0.17(1) \hspace{5mm} \mbox{for}\ K=0.1925, \ L=12.
\end{eqnarray}
 The string tension in lattice units is decreasing when the critical
 line is approached, as it should be.
 This is mainly caused by the renormalization of the gauge coupling due
 to virtual gluino loop effects which are manifested by decreasing
 lattice spacing $a$.
 From a comparison of the $L=8$ and $L=12$ results one sees that finite
 size effects still appear to be sizable.
 This has to be expected because we have for the spatial lattice
 extension $L=12a$ the result $L\sqrt{\sigma} \simeq 2.1$.
 In QCD with $\sqrt{\sigma} \simeq 0.45\, GeV$ this would correspond
 to $L \simeq 1\, fm$.
 Although we are dealing with a different theory where finite size
 effects as a function of $L\sqrt{\sigma}$ are different, for a first
 orientation this estimate should be good enough.

 For the ratio of the scalar glueball mass $m(0^+)$, to be discussed
 below, and the square root of the string tension the results are:
\begin{eqnarray}\label{eq506}
m(0^+) / \sqrt{\sigma} &=& 3.4(7) \hspace{3mm}
\mbox{for}\ K=0.1900, \ L=8,                       \nonumber \\
m(0^+) / \sqrt{\sigma} &=& 3.0(4) \hspace{3mm}
\mbox{for}\ K=0.1925, \ L=8,                       \nonumber \\
m(0^+) / \sqrt{\sigma} &=& 3.1(7) \hspace{3mm}
\mbox{for}\ K=0.1925, \ L=12.
\end{eqnarray}
 The uncertainties are not very small, but the numbers are consistent
 with a constant independent of $K$ in this range.

\subsection{Supersymmetry multiplets?}\label{sec5.2}
 The non-vanishing string tension implies that the Yang-Mills theory
 with gluinos is confining.
 Therefore the asymptotic states are colour singlets, similarly to
 hadrons in QCD.
 The structure of the light hadron spectrum is closest to the
 (theoretical) case of QCD with a single flavour of quarks where the
 chiral symmetry is broken by the anomaly.

 Since both gluons and gluinos transform according to the adjoint
 (for SU(2) gauge group triplet) representation of the colour group, one
 can construct colour singlet interpolating fields from any number of
 gluons and gluinos if their total number is at least two.
 Experience in QCD suggests that the lightest states can be well
 represented by interpolating fields built out of a small number of
 constituents.
 Simple examples are the {\em glueballs} known from pure Yang-Mills
 theory and {\em gluinoballs} which are composite states made out of
 gluinos.
 Mixed {\em gluino-glueball} states can be composed of any number of
 gluons and any number of gluinos, in the simplest case just one of
 both.

 In general, one has to keep in mind that the classification of states
 by some interpolating fields has only a limited validity, because
 this is a strongly interacting theory where many interpolating fields
 can have important projections on the same state.
 Taking just the simplest colour singlets can, however, give a good
 qualitative description.

 In the supersymmetric limit at zero gluino mass $m_{\tilde{g}}=0$ the
 hadronic states should occur in supermultiplets.
 This restricts the choice of simple interpolating field combinations
 and leads to low energy effective actions in terms of them 
 \cite{VENYAN,FAGASCH}.
 For non-zero gluino mass the supersymmetry is softly broken and
 the hadron masses are supposed to be analytic functions of
 $m_{\tilde{g}}$.
 The linear terms of a Taylor expansion in $m_{\tilde{g}}$ are often
 determined by the symmetries of the low energy effective actions
 \cite{EVHSSC}.

 The effective action of Veneziano and Yankielowicz \cite{VENYAN} in
 eq.~(\ref{eq123}) describes a chiral supermultiplet consisting of the
 $0^-$ gluinoball $\tilde{\eta}$, the $0^+$ gluinoball $\tilde{f}_0$ and
 a spin $\half$ gluino-glueball $\tilde{\chi}$.
 There is, however, no a priori reason to assume that glueball states
 are heavier than the members of this supermultiplet.
 Therefore Farrar, Gabadadze and Schwetz \cite{FAGASCH} proposed an
 effective action which includes an additional chiral supermultiplet.
 This consists of a $0^+$ glueball, a $0^-$ glueball and another
 gluino-glueball.
 The effective action allows mass mixing between the members of the two
 supermultiplets.

 The spectrum of SYM theory with SU(2) gauge group has been studied
 in numerical simulations by the DESY-M\"unster Collaboration
 \cite{SPECTRUM,PISA,KIRCHNER}.
 The glueball states as well as the methods to compute their masses
 in numerical Monte Carlo simulations are well known from pure gauge
 theory (see \cite{GLUEBALL}).
 The obtained masses for the $0^+$ glueball in lattice units are
\begin{eqnarray}\label{eq507}
a m(0^+) &=& 0.95(10) \hspace{3mm} \mbox{for}\ K=0.1800, \ L=6,
\nonumber \\
a m(0^+) &=& 0.85(6) \hspace{5mm}  \mbox{for}\ K=0.1850, \ L=6,
\nonumber \\
a m(0^+) &=& 0.75(6) \hspace{5mm}  \mbox{for}\ K=0.1900, \ L=8,
\nonumber \\
a m(0^+) &=& 0.63(5) \hspace{5mm}  \mbox{for}\ K=0.1925, \ L=8,
\nonumber \\
a m(0^+) &=& 0.53(10) \hspace{3mm} \mbox{for}\ K=0.1925, \ L=12.
\end{eqnarray}
 As before, $L$ denotes the spatial extension of the lattice.
 (The time extension has always been $T=2L$.)

 In addition to the $J^P=0^+$ glueball one can also search for the
 pseudoscalar $0^-$ glueball.
 The masses in lattice units turned out to be
\begin{eqnarray}\label{eq508}
a m(0^-) &=& 1.5(3) \hspace{7mm} \mbox{for}\ K=0.1850, \ L=6,
\nonumber \\
a m(0^-) &=& 1.45(10) \hspace{3mm} \mbox{for}\ K=0.1900, \ L=6,
\nonumber \\
a m(0^-) &=& 1.3(1) \hspace{7mm} \mbox{for}\ K=0.1925, \ L=6,
\nonumber \\
a m(0^-) &=& 1.1(1) \hspace{7mm} \mbox{for}\ K=0.1925, \ L=8.
\end{eqnarray}
 The pseudoscalar glueball appears to be roughly twice as heavy as the
 scalar one.
 This is similar to pure SU(2) gauge theory, where
 $m(0^-)/m(0^+) = 1.8(2)$ \cite{GLUEBALL}.

 As discussed above, besides the glueballs made out of gluons one also
 has to consider gluinoballs.
 Examples of simple colourless composite fields can be constructed from
 two gluino fields: $\overline{\lambda}_x\gamma_5\lambda_x$ and
 $\overline{\lambda}_x\lambda_x$.
 These correspond to the gluinoball states mentioned above:
 $\tilde{\eta}$ and $\tilde{f}_0$, respectively, which are described by
 the effective action (\ref{eq123}).

 The masses of $\tilde{\eta}$ and $\tilde{f}_0$ can be extracted
 from correlation functions as
\be\label{eq509}
\Gamma_{\tilde {g} \tilde {g} } (x,y) =
\langle {\rm Tr_{sc}}\, \{ \Gamma Q^{-1}_{xx} \} 
{\rm Tr_{sc}}\, \{ \Gamma Q^{-1}_{yy} \} -
2\, {\rm Tr_{sc}}\, \{ \Gamma Q^{-1}_{xy} \Gamma Q^{-1}_{yx} \} \rangle
\ ,
\ee
 where $\Gamma \in \{1,\gamma_5 \}$ and $Q^{-1}$ is the gluino
 propagator.
 Note that the factor of two is originating from the Majorana character
 of the gluinos.
 In analogy with the flavour singlet meson in QCD the correlator in
 (\ref{eq509}) consists of a connected and a disconnected part.
 The disconnected part can be calculated using the volume source
 technique \cite{KUFUMIOKUK}.

 In case of the $\tilde{f}_0$ particle the disconnected and the connected
 parts are of the same order of magnitude whereas $\tilde{\eta}$ is
 dominated by the connected part.
 The disconnected part has a much worse signal to noise ratio than the
 connected one.
 This leads to a larger error on the $\tilde{f}_0$ mass as compared to
 the $\tilde{\eta}$ mass.

 In the low energy effective action of Farrar, Gabadadze and Schwetz
 \cite{FAGASCH} there is a possible non-zero mixing between the
 states in two light supermultiplets.
 In particular there can be mixing of the $\tilde{f}_0$ gluinoball and
 the $0^+$ glueball which have identical quantum numbers.
 The numerical simulations show that this mixing is small
 \cite{SPECTRUM,KIRCHNER}.

\begin{figure}[htb]
\vspace*{0.2cm}
\begin{center}
\epsfig{file=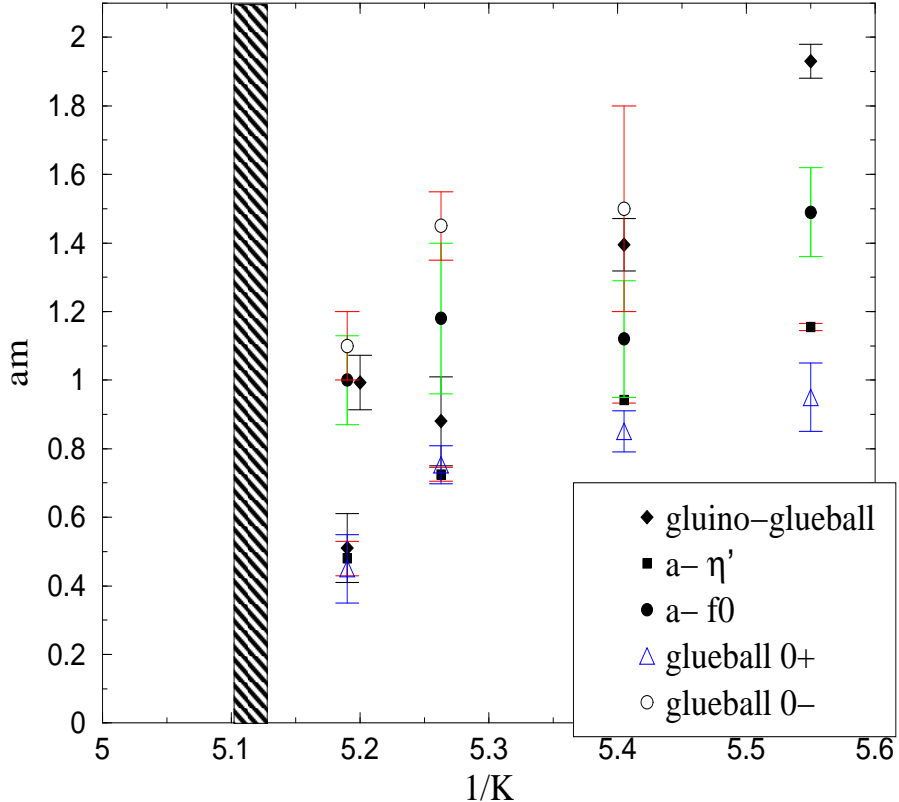,
        width=12cm,height=11cm,angle=0}
\end{center}
\vspace*{-1.2cm}
\begin{center}
\parbox{12cm}{\caption{\label{fig08}\em
 The lightest bound state masses in lattice units as function of the
 bare gluino mass parameter $1/K$ for fixed $\beta=2.3$.
 The shaded area at $K=0.1955(5)$ is where zero gluino mass and
 supersymmetry are expected.
 In this figure the notations of \protect\cite{SPECTRUM,PISA,KIRCHNER}
 are used: $a$-$\eta^\prime\equiv\tilde{\eta}$ and
 $a$-$f_0\equiv\tilde{f}_0$.}}
\end{center}
\vspace*{-0.5cm}
\end{figure}

 The low mass supermultiplets containing $J^P=0^-,0^+$ states have to
 be completed by a spin $\half$ state.
 The corresponding composite fields can constructed from the gluino
 field and the field strength tensor of gluons.
 The simplest example is \cite{DGHV,KIRCHNER}:
\be\label{eq510}
\Phi \equiv \sigma_{\mu\nu}\, {\rm Tr\,} \left[ F_{\mu\nu}\lambda \right]
\ .
\ee
 On the lattice one has to use, of course, an appropriate construction
 for $F_{\mu\nu}$ out of the link variables $U_{x\mu}$.
 Composite fields with the same quantum numbers can also be built up
 from three gluino fields \cite{PISA,KIRCHNER}.
 A summary of the results of the DESY-M\"unster Collaboration about
 the spectrum of light composite states is shown in figure~\ref{fig08}.

 As the figure shows, the low mass states appear to build two groups
 which may correspond to the effective action \cite{FAGASCH}.
 For a firm conclusion more numerical work is needed in larger volumes
 and for several values of the gauge coupling $\beta$ allowing for
 a continuum extrapolation.

\section{Supersymmetric Ward-Takahashi identities}\label{sec6}
 An important feature of lattice regularization is that some symmetries
 are broken for non-zero lattice spacing and are expected to be
 recovered in the continuum limit.
 The details of the lattice formulation, which may also influence the
 degree of symmetry breaking, are not relevant in the continuum limit
 because of the {\em universality of critical points}.
 A basic set of symmetries broken by the lattice and restored in the
 continuum limit is the (Euclidean) Lorentz symmetry including rotations
 and translations.
 It is clear that on any regular lattice these symmetries are always
 broken.
 Internal symmetries as, for instance, global chiral symmetry are
 sometimes broken sometimes conserved on the lattice, depending on the
 actual formulation.
 From the point of view of symmetry realization supersymmetry is
 expected to behave similarly to the Lorentz symmetry: at finite lattice
 spacing it is broken but it becomes restored in the continuum limit.
 This similarity is quite natural since there is an intimate relation
 of supersymmetry to the Lorentz symmetries of space-time shown, for
 instance, by the fact that the anticommutators of the supersymmetry
 charges give translations.

 In the framework of quantum field theory the symmetries can be
 exploited by the corresponding Ward-Takahashi (WT) identities.
 Well studied examples are the global axial (chiral) symmetries in
 lattice QCD \cite{BMMRT}.
 In case of SYM theory this way of realizing supersymmetry has been
 first considered by Curci and Veneziano \cite{CURVEN}.
 At zero gaugino mass both supersymmetry and anomalous chiral symmetry
 has to be manifested in the corresponding WT identities.

 Before going to the lattice formulation let us describe the WT
 identities of supersymmetry in the continuum at a somewhat formal,
 non-rigorous level (for a more rigorous treatment see \cite{RUSI}).
 The corresponding {\em supercurrent} $S_\mu$ has been introduced in
 (\ref{eq105}):
\be\label{eq601}
S_\mu(x) \equiv 
-F^a_{\rho\tau}(x)\sigma_{\rho\tau}\gamma_\mu \lambda^a(x) =
-\frac{2i}{g} \sigma_{\rho\tau}\gamma_\mu \,
{\rm Tr\,}\left(F_{\rho\tau}(x)\lambda(x)\right) \ ,
 \ee
 where the field strength tensor matrix $F_{\mu\nu}(x)$ is defined in
 (\ref{eq107}) and the gaugino field matrix is
 $\lambda(x) \equiv \lambda^a(x) T_a$.
 Since the regularization breaks supersymmetry the simple conservation
 low $\partial_\mu S_\mu(x)=0$ can only be recovered in the process of
 renormalization if the possible operator mixings are properly taken
 into account.
 The analysis of the operator mixings \cite{WTI} leads to the following
 conjecture about the form of SUSY-WT identities:
\be\label{eq602}
\langle Z_S\partial_\mu S_\mu(x) {\cal O}(y) +
Z_T\partial_\mu T_\mu(x) {\cal O}(y) \rangle =
(m_0-\bar{m})\langle D_S(x) {\cal O}(y) \rangle \ .
 \ee
 Here ${\cal O}(y)$ is any gauge invariant local operator at point $y$
 and $T_\mu(x)$ is another dimension $\frac{7}{2}$ operator besides
 $S_\mu(x)$, namely
\be\label{eq603}
T_\mu(x) = \frac{2i}{g}\gamma_\nu \, 
{\rm Tr\,}\left(F_{\mu\nu}(x)\lambda(x)\right) \ .
 \ee
 The operator $D_S(x)$ appearing on the right hand side of (\ref{eq602})
 is defined as
\be\label{eq604}
D_S(x) = \frac{2i}{g}\sigma_{\mu\nu} \, 
{\rm Tr\,}\left(F_{\mu\nu}(x)\lambda(x)\right) \ .
 \ee
 $Z_S$ and $Z_T$ are multiplicative renormalization factors, $m_0$ is
 the bare gaugino mass and $\bar{m}$ represents the additive
 renormalization of the bare mass.
 The form of (\ref{eq602}) is valid only for $x \ne y$ because
 ``contact terms'' have been omitted.
 The consequence of (\ref{eq602}) is that the (renormalized) gaugino
 mass vanishes if $m_0-\bar{m}=0$ and the renormalized supercurrent
 can be defined as
\be\label{eq605}
\hat{S}_\mu(x) \equiv Z_S S_\mu(x) + Z_T T_\mu(x) \ .
 \ee
%

\subsection{Lattice formulation}\label{sec6.1}
 The arguments leading to the SUSY-WT identity (\ref{eq602}) can be
 followed step by step in lattice regularization.
 First one has to define the supersymmetry transformations.
 Since SUSY is broken on the lattice the lattice action will not be
 invariant with respect to these transformations.
 There is a freedom in the definition which is only restricted by the
 requirement that in the continuum limit the infinitesimal
 transformations in (\ref{eq103}) have to be reproduced.
 The supersymmetry breaking terms, which should vanish as ${\cal O}(a)$
 in the continuum limit $a \to 0$, can be minimized by an appropriate
 choice of the irrelevant parts.
 An important requirement is to maintain as many discrete lattice
 symmetries as possible.
 In particular the parity ${\cal P}$, time reversal ${\cal T}$ and
 charge conjugation ${\cal C}$ transformations should commute with
 the lattice SUSY transformations.

 A simple choice fulfilling these requirement has been introduced in
 \cite{TANIG}.
 For these definitions let us change the conventions in the lattice
 action for gauginos (\ref{eq216}) by changing the normalization of the
 gaugino field according to
\be\label{eq606}
\lambda_x \to \sqrt{m_0+4}\; \lambda_x
 \ee
 and introduce the bare lattice mass $m_0$ by
\be\label{eq607}
m_0 \equiv \frac{1}{2K}-4 \ , \hspace{3em}
K = \frac{1}{2(m_0+4)} \ .
 \ee
 Using the gaugino field matrix $\lambda_x \equiv \lambda^a_x T_a$ we
 have instead of (\ref{eq216}) the fermionic action
\be\label{eq608}
S_f = \sum_x {\rm Tr\,} \LCB (m_0+4)\overline{\lambda}_x\lambda_x
-\half \sum_{\mu=1}^4 \left[
\overline{\lambda}_{x+\hat{\mu}} U_{x\mu}(1+\gamma_\mu)
\lambda_x U^\dagger_{x\mu} +
\overline{\lambda}_x U^\dagger_{x\mu} (1-\gamma_\mu)
\lambda_{x+\hat{\mu}} U_{x\mu} \right] \RCB \ .
\ee
 (Note that compared to ref.~\cite{TANIG} $U_{x\mu}$ is replaced
 here by $U^\dagger_{x\mu}$.)
 This notation is close to the continuum conventions whereas
 (\ref{eq216}) is practical for numerical simulations.
 For the definition of the SUSY transformations we need an appropriately
 defined field strength tensor on the lattice which we denote by
 $P_{\mu\nu}(x)$:
\begin{eqnarray}\label{eq609}
P_{\mu\nu}(x) & \equiv & \frac{1}{8ig} \sum_{i=1}^4 
\left[ U_{i\mu\nu}(x) - U^\dagger_{i\mu\nu}(x) \right] \ ,
\nonumber \\
U_{1\mu\nu}(x) & \equiv & U^\dagger_{x\nu} U^\dagger_{x+\hat{\nu},\mu}
U_{x+\hat{\mu},\nu} U_{x\mu} \ ,
\nonumber \\
U_{2\mu\nu}(x) & \equiv & U^\dagger_{x\mu} U_{x-\hat{\nu}+\hat{\mu},\nu}
U_{x-\hat{\nu},\mu} U^\dagger_{x-\hat{\nu},\nu} \ ,
\nonumber \\
U_{3\mu\nu}(x) & \equiv & U_{x-\hat{\nu},\nu} U_{x-\hat{\nu}-\hat{\mu},\mu}
U^\dagger_{x-\hat{\nu}-\hat{\mu},\nu} U^\dagger_{x-\hat{\mu},\mu} \ ,
\nonumber \\
U_{4\mu\nu}(x) & \equiv & U_{x-\hat{\mu},\mu} U^\dagger_{x-\hat{\mu},\nu}
U^\dagger_{x+\hat{\nu}-\hat{\mu},\mu} U_{x\nu} \ .
\end{eqnarray}
 This transforms under parity and time reversal in the same way as
 $F_{\mu\nu}$ does in the continuum.
 Using $P_{\mu\nu}(x)$ one can define the infinitesimal SUSY
 transformations on the lattice by
\begin{eqnarray}\label{eq610}
\delta_\xi U_{x\mu} &=& -\frac{ig}{2} \left[
\overline{\xi} \gamma_\mu U_{x\mu} \lambda_x +
\overline{\xi} \gamma_\mu \lambda_{x+\hat{\mu}} U_{x\mu}
\right] \ ,
\nonumber \\
\delta_\xi U^\dagger_{x\mu} &=& \frac{ig}{2} \left[
\overline{\xi} \gamma_\mu \lambda_x U^\dagger_{x\mu}  +
\overline{\xi} \gamma_\mu U^\dagger_{x\mu} \lambda_{x+\hat{\mu}}
\right] \ ,
\nonumber \\
\delta_\xi \lambda_x &=& \half\sigma_{\mu\nu} \xi P_{\mu\nu}(x) \ ,
\nonumber \\
\delta_\xi \overline{\lambda}_x &=& -\half \overline{\xi}
\sigma_{\mu\nu} P_{\mu\nu}(x) \ .
\end{eqnarray}
 Here $\xi$ and $\overline{\xi}$ are Grassmannian parameters satisfying
 a Majorana condition like (\ref{eq102}).

 After defining the lattice SUSY transformations (\ref{eq610}) one can
 derive the SUSY WT-identities by a standard procedure \cite{BMMRT,MM}.
 The corresponding lattice supercurrent $S_\mu(x)$ can be identified
 as
\be\label{eq611}
S_\mu(x) = -\half \sum_{\rho\sigma} \sigma_{\rho\sigma} \gamma_\mu
{\rm \,Tr\,} \left[ 
P_{\rho\sigma}(x) U^\dagger_{x\mu} \lambda_{x+\hat{\mu}} U_{x\mu} +
P_{\rho\sigma}(x+\hat{\mu}) U_{x\mu} \lambda_x U^\dagger_{x\mu}
\right] \ .
\ee
 The spinorial density multiplying the gaugino mass turns out to be
\be\label{eq612}
D_S(x) = \sum_{\rho\sigma} \sigma_{\rho\sigma} {\rm \,Tr\,} \left[
P_{\rho\sigma}(x) \lambda_x \right] \ .
\ee
 The other supercurrent $T_\mu(x)$ which is mixed to $S_\mu(x)$ can be
 defined as
\be\label{eq613}
T_\mu(x) = \sum_\nu \gamma_\nu {\rm \,Tr\,} \left[ 
P_{\mu\nu}(x) U^\dagger_{x\mu} \lambda_{x+\hat{\mu}} U_{x\mu} + 
P_{\mu\nu}(x+\hat{\mu}) U_{x\mu} \lambda_x U^\dagger_{x\mu}
\right] \ .
\ee
 In the continuum limit, apart from differences in normalization
 conventions, these supercurrents go over to their continuum
 conterparts in (\ref{eq601})-(\ref{eq604}).
 
 Using these definitions the SUSY WT-identity can be written as
 \cite{CURVEN,DGHV,TANIG}:
\be\label{eq614}
Z_S \langle \nabla^b_\mu S_\mu(x) {\cal O}(y) \rangle +
Z_T \langle \nabla^b_\mu T_\mu(x) {\cal O}(y) \rangle =
(m_0-\bar{m})\langle D_S(x) {\cal O}(y) \rangle + {\cal O}(a) \ .
 \ee
 Here the lattice (backward) derivative is defined as
 $\nabla^b_\mu f(x) \equiv f(x)-f(x-\hat{\mu})$.
 The gauge invariant function ${\cal O}(y)$ at point $y$ is assumed to
 be sufficiently far away from $x$ in such a way that no common
 points with the expressions defined at $x$ occur.
 In this way additional {\em contact terms} are avoided.

 The SUSY WT-identity as it stands in (\ref{eq614}) is only valid for
 gauge invariant functions ${\cal O}(y)$ of lattice fields.
 In case of gauge non-invariance several additional terms appear.
 Examples of such anomalous terms emerge, for instance, in lattice
 perturbation theory calculations \cite{TANIG,SUSYWTP}.
 The reason behind these complications is the conflict of gauge
 fixing with supersymmetry.

 As already remarked above, the form of the lattice currents in
 (\ref{eq611})-(\ref{eq613}) is not unique because terms of order
 ${\cal O}(a)$ can be added.
 Locality and simplicity of the expressions is always welcome.
 This leads to the alternative definitions of the ``local''
 supercurrents
\be\label{eq615}
S^l_\mu(x) = -\sum_{\rho\sigma} \sigma_{\rho\sigma} \gamma_\mu
{\rm \,Tr\,} \left[ P_{\rho\sigma}(x) \lambda_x \right]
\ee
 and
\be\label{eq616}
T^l_\mu(x) = 2\sum_\nu \gamma_\nu {\rm \,Tr\,} \left[ 
P_{\mu\nu}(x) \lambda_x \right] \ .
\ee
 In contrast to these we can call the supercurrents in
 (\ref{eq611}) and (\ref{eq613}) as ``point-split'':
\be\label{eq617}
S^{ps}_\mu(x) \equiv S_\mu(x) \ , 
\hspace{3em}
T^{ps}_\mu(x) \equiv T_\mu(x) \ .
\ee

 In addition to the ambiguity of supercurrents the lattice approximation
 of the derivatives can also be varied. For instance, for the local
 supercurrents in (\ref{eq615}) and (\ref{eq616}) the backward
 difference $\nabla^b\mu$ may be replaced by a ``symmetric difference''
 $\nabla^s_\mu f(x) \equiv \half (f(x+\hat{\mu})-f(x-\hat{\mu}))$.
 All these variations are irrelevant in the continuum limit but for
 non-zero lattice spacing $a \ne 0$ the ${\cal O}(a)$ lattice artefacts
 in (\ref{eq614}) can be rather different.

\begin{figure}[htb]
\vspace*{0.2cm}
\begin{center}
\epsfig{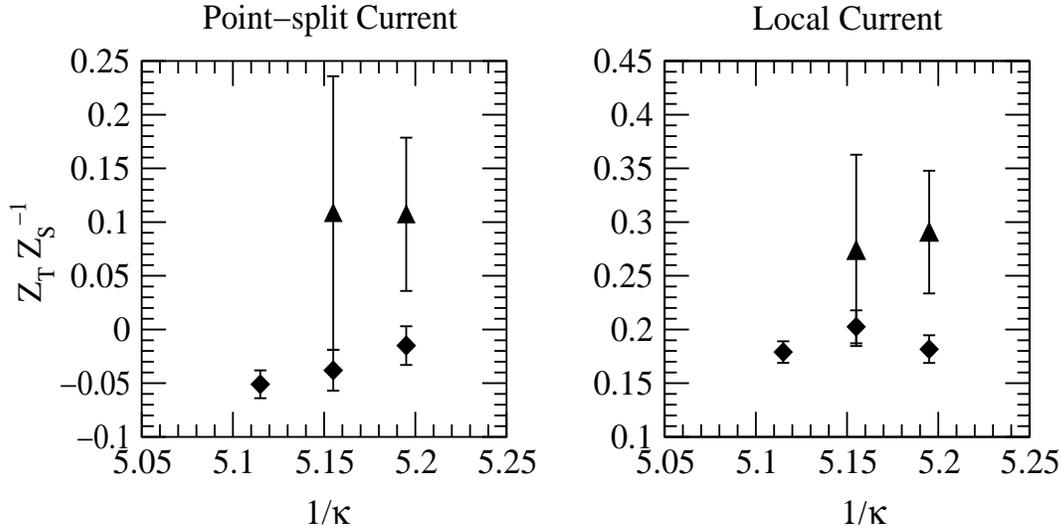}
\end{center}
\vspace*{-0.7cm}
\begin{center}
\parbox{12cm}{\caption{\label{fig09}\em
 Results for the supercurrent renormalization factors from
 \protect\cite{WTI}.
 On the horizontal axis the inverse hopping parameter values are shown
 ($1/\kappa \equiv 1/K$).}}
\end{center}
\vspace*{-0.5cm}
\end{figure}

\begin{figure}[htb]
\vspace*{0.2cm}
\begin{center}
\epsfig{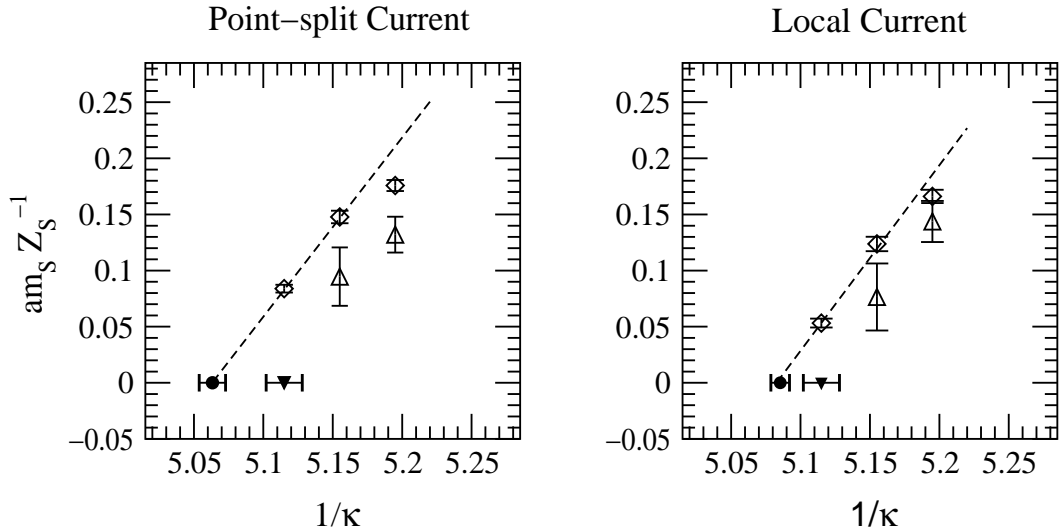}
\end{center}
\vspace*{-1.2cm}
\begin{center}
\parbox{12cm}{\caption{\label{fig10}\em
 Results for the renormalized gaugino mass from \protect\cite{WTI}.
 On the horizontal axis the invers hopping parameter values are shown
 ($1/\kappa \equiv 1/K$).
 The dashed lines are linear extrapolations for vanishing gaugino
 mass.}}
\end{center}
\vspace*{-0.5cm}
\end{figure}

\subsection{Numerical results}\label{sec6.2}
 The DESY-M\"unster-Roma Collaboration studied the SUSY WT-identities
 in numerical simulations \cite{WTIB,WTI}.
 Omitting ${\cal O}(a)$ terms and dividing by $Z_S$ the SUSY WT-identity
 in (\ref{eq614}) can be brought to the form \cite{DGHV}
\be\label{eq618}
\langle \nabla^b_\mu S_\mu(x) {\cal O}(y) \rangle +
\frac{Z_T}{Z_S} \langle \nabla^b_\mu T_\mu(x)
{\cal O}(y) \rangle = \frac{(m_0-\bar{m})}{Z_S}
\langle D_S(x) {\cal O}(y) \rangle \ .
\ee
 This has to be valid for every gauge invariant function ${\cal O}(y)$
 which is defined in a point $y$ in such a way that there are no
 common points with the functions defined at $x$.
 Considering (\ref{eq618}) with different ${\cal O}(y)$ a system of
 linear equations is obtained for the two unknowns
\be\label{eq619}
z_{TS} \equiv \frac{Z_T}{Z_S} \ , \hspace{3em}
z_{MS} \equiv \frac{(m_0-\bar{m})}{Z_S} \ .
\ee
 (Note that $z_{MS}$ is proportional to the renormalized gaugino mass
 $m_R \equiv (m_0-\bar{m})/Z_m$.)
 Since there are very many different possible ${\cal O}(y)$'s, the
 expectation that (\ref{eq618}) has, in the continuum limit, a unique
 solution pair $(z_{TS},z_{MS})$ is highly non-trivial.
 Its numerical investigation can strongly support (or perhaps
 contradict) our expectations about the realization of supersymmetry in
 SYM theory.

 For obtaining non-zero expectation values the gauge invariant functions
 ${\cal O}(y)$ appearing in (\ref{eq619}) have to be chosen
 appropriately.
 Roughly speaking they should have quantum numbers similar to the
 spinorial density $D_S(x)$.
 The intermediate states with important contributions are composite
 states of a gluino (= gaugino) and a gluon as, for instance, the
 gluino-glueball $\tilde{\chi}$ investigated in section \ref{sec5.2}.
 In order to obtain a good signal to noise ratio one can sum over
 three $x$-coordinates $(x_1,x_2,x_3)$ keeping the fourth
 time-coordinate fixed.
 This projects out intermediate states with zero spatial momentum.
 One can also apply {\em optimized smearing techniques} in the timeslice
 containing $y$ similarly to correlators for obtaining the masses of
 gluino-glueballs.

\begin{table}[bh]
\begin{center}
\parbox{15cm}{\caption{\label{tab02}
 The values of $r_0/a$ for the sets of configurations considered in
 \protect{\cite{WTI}}.
 The ratio $L/r_0$ is also shown, where $L$ is the spatial lattice
 size.}}
\end{center}
\begin{center}
\begin{tabular}{lll}
\hline
\multicolumn{1}{c}{$\kappa$ }  &
\multicolumn{1}{c}{$r_0/a$}  &
\multicolumn{1}{c}{$L/r_0$}  \\
\hline
0.1925 & 6.71(19) & 1.79(5)  \\
0.194 & 7.37(30) & 1.63(7)  \\
0.1955 & 7.98(48) & 1.50(9)  \\
\hline
\end{tabular}
\end{center}
\end{table}

 The numerical simulations of the DESY-M\"unster-Roma Collaboration
 \cite{WTI} has been performed on a $12^3 \cdot 24$ lattice at
 $\beta=2.3$ for three different values of the bare gaugino mass
 given by $K=0.1925,\; 0.1940,\; 0.1950$.
 These parameters are well suited for a first study because they
 are reasonably close to the continuum limit and since the WT-identities
 hold in any volume no finite volume effects are disturbing.
 The values of the scale parameter $r_0$ \cite{SOMMER}, which give
 information about the physical volume size, are shown in
 table~\ref{tab02}.
 A summary of the resulting fits to $z_{TS}$ and $z_{MS}$ is contained
 in figures~\ref{fig09} and \ref{fig10}.
 (For more details see \cite{WTI}.)

 The renormalization factors $Z_S$ and $Z_T$ depend on the definition
 of the lattice supercurrents.
 As figure~\ref{fig09} shows, the admixture of $T_\mu(x)$ is negligible
 for the point-split supercurrent $S^{ps}_\mu(x)$ but non-zero for the
 local supercurrent $S^l_\mu(x)$.
 $z_{TS}$ is within errors independent of the bare mass given by $K$.
 $z_{MS}$ is proportional to the renormalized mass and tends to zero
 if $K$ approaches the critical value $K_{cr}$.

 These results are consistent with expectations, in particular with the
 realization of supersymmetry in a Yang-Mills theory with massless
 Majorana fermion in the adjoint representation.
 For a stronger conclusion the continuum limit $\beta\to\infty$ has
 to be investigated in future simulations.

\section{Outlook}\label{sec7}
 The numerical Monte Carlo simulation of supersymmetric Yang-Mills
 theory is feasible with presently available computational resources
 and with known algorithms.
 The total computational effort of the DESY-M\"unster-Roma Collaboration
 has been of the order of 10 Gflops $\cdot$ year
 $\simeq 3 \cdot 10^{17}$ floating point operations.
 This is much less than the amount of computer time devoted up to now
 to QCD.
 In fact, to simulate SYM theory is substantially easier than QCD
 because the ``flavour number'' of fermions $N_f=\half$ is smaller.
 In addition, in contrast to QCD, in SYM there are no almost massless
 pseudoscalar mesons expected.
 The easiness and the considerable theoretical interest make the Monte
 Carlo simulation of SYM theory a promising future subject.

 The next steps in improving the present simulations are obvious:
\begin{itemize}
\item
 clarification of the nature of the transition at vanishing gaugino mass
 by studying its behaviour on larger lattice volumes;
\item
 investigation of the spectrum of bound states in supermultiplets in
 sufficiently large volumes and closer to the continuum limit;
\item
 study of the continuum limit of the supersymmetric Ward-Takahashi
 identities.
\end{itemize}

 Once these basic questions are sufficiently clarified one can certainly
 find other more detailed questions which will help understanding many
 aspects of the dynamics of supersymmetric gauge field theories.

\vspace{10mm}                                                    
{\large\bf Acknowledgements} 

\vspace{5mm}\noindent
 The numerical simulation results summarized in this review have been
 obtained due to the enthusiastic work done by the members of the
 DESY-M\"unster-Roma Collaboration.
 I thank them for valuable discussions and comments.
 The Monte Carlo runs have been performed on the CRAY-T3E
 supercomputers at NIC J\"ulich.
 It is a pleasure to thank the staff at NIC for their kind support.




\newpage
\begin{thebibliography}{99}
%
\bibitem{SEIWIT}
N. Seiberg and E. Witten,
{\em Nucl. Phys.} {\bf B426}, 19 (1994); 
ERRATUM {\em ibid.} {\bf B430}, 485 (1994).
%
\bibitem{MAJORANA}
H. Nicolai,
{\em Nucl. Phys.} {\bf B140}, 294 (1978);\\
P. van Nieuwenhuizen, A. Waldron,
{\em Phys. Lett.} {\bf B389}, 29 (1996).
%
\bibitem{SUSY}
J. Bagger, J. Wess,
{\em Supersymmetry and Supergravity},
Princeton University Press, 1983;      \\
P. Fayet, S. Ferrara,
{\em Phys. Rep.} {\bf 32}, 249 (1977);  \\
M.F. Sohnius,
{\em Phys. Rep.} {\bf 128}, 39 (1985).
%
\bibitem{RUSI}
C. Rupp, K. Sibold,
hep-th/0101165.
%
\bibitem{PISASU3}
A. Feo et al. (DESY-M\"unster Collaboration),
{\em Nucl. Phys. Proc. Suppl.} {\bf 83}, 661 (2000).
%
\bibitem{AFDISE}
I. Affleck, M. Dine, N. Seiberg,
{\em Phys. Rev. Lett.} {\bf 51}, 1026 (1983);
{\em Nucl. Phys.} {\bf B241}, 493 (1984).
%
\bibitem{NOSHVAZA}
V.A. Novikov, M.A. Shifman, A.I. Vainshtein, V.I. Zakharov,
{\em Nucl. Phys.} {\bf B260}, 157 (1985).
%
\bibitem{SHIVAI}
M.A. Shifman, A.I. Vainshtein,
{\em Nucl. Phys.} {\bf B296}, 445 (1988).
%
\bibitem{FINPOU}
D. Finnell, P. Pouliot,
{\em Nucl. Phys.} {\bf B453}, 225 (1995).
%
\bibitem{NSVZ}
V.A. Novikov, M.A. Shifman, A.I. Vainshtein, V.I. Zakharov,
{\em Nucl. Phys.} {\bf B229}, 407 (1983).
%
\bibitem{ROSVEN}
G.C. Rossi, G. Veneziano,
{\em Phys. Lett.} {\bf B138}, 195 (1984).
%
\bibitem{AMROVE}
D. Amati, G.C. Rossi, G. Veneziano,
{\em Nucl. Phys.} {\bf B249}, 1 (1985).
%
\bibitem{AKMRV}
D. Amati, K. Konishi, Y. Meurice, G.C. Rossi,G. Veneziano, 
{\em Phys. Rep.} {\bf 162}, 169 (1988).
%
\bibitem{HOKHLEMA}
T.J. Hollowood, V.V. Khoze, W. Lee, M.P. Mattis,
{\em Nucl. Phys.} {\bf B570}, 241 (2000).
%
\bibitem{KOVSHI}
A. Kovner, M. Shifman,
{\em Phys. Rev.} {\bf D56}, 2396 (1997).
%
\bibitem{WEISZ}
P. Weisz,
{\em Phys. Lett.} {\bf 100B}, 331 (1981) and private communication.
%
\bibitem{WITTEN}
E. Witten,
{\em Nucl. Phys.} {\bf B202}, 253 (1982).
%
\bibitem{VENYAN}
G. Veneziano, S. Yankielowicz,
{\it Phys. Lett.} {\bf B113}, 231 (1982).
%
\bibitem{FERZUM}
S. Ferrara, B. Zumino,
{\it Nucl. Phys.} {\bf B87} 207 (1975).
%
\bibitem{EVHSSC}
N. Evans, S.D.H. Hsu, M. Schwetz, 
hep-th/9707260.
%
\bibitem{FAGASCH}
G.R. Farrar, G. Gabadadze, M. Schwetz,
{\em Phys. Rev.} {\bf D60}, 035002 (1999).
%
\bibitem{KOSHSM}
A. Kovner, M. Shifman, A. Smilga, 
{\em Phys. Rev.} {\bf D56}, 7978 (1997).
%
\bibitem{GLUEBALL}
C. Michael, M. Teper,
{\em Phys. Lett.} {\bf B199}, 95 (1987); \\
H. Chen, J. Sexton, A. Vaccarino, D. Weingarten,
{\em Nucl. Phys. Proc. Suppl.} {\bf 34}, 357 (1994); \\
M.J. Teper, in
{\em Confinement, Duality and Non-Perturbative Aspects of QCD,}
NATO Advanced Study Institute, Cambridge 1997 and hep-lat/9711011.
%
\bibitem{JACJON}
I. Jack, D.R.T. Jones,
in {\em Perspectives on supersymmetry,} ed. G.L. Kane;
hep-ph/9705417.
%
\bibitem{CASHAM}
A. Casher, Y. Shamir,
hep-th/9908074.
%
\bibitem{MM}
I. Montvay, G. M\"unster,
{\em Quantum Fields on a Lattice,} Cambridge University Press,
1994.
%
\bibitem{DGHV}
A. Donini, M. Guagnelli, P. Hernandez, A. Vladikas,
{\em Nucl. Phys.} {\bf B523}, 529 (1998).
%
\bibitem{BIETEN}
W. Bietenholz,
{\em Mod. Phys. Lett.} {\bf A14}, 51 (1999).
%
\bibitem{CATKAR}
S. Catterall, S. Karamov,
hep-lat/010071.
%
\bibitem{KAPSCH}
D.B. Kaplan, M. Schmaltz,
{\em Chin. J. Phys.} {\bf 38}, 543 (2000).
%
\bibitem{NEUBERGER}
H. Neuberger,
{\em Phys. Rev.} {\bf D57}, 5417 (1998).
%
\bibitem{HOIZNI}
T. Hotta, T. Izubuchi, J. Nishimura,
{\em Mod. Phys. Lett.} {\bf A13}, 1667 (1998).
%
\bibitem{GINWIL}
P.H. Ginsparg, K.G. Wilson,
{\em Phys. Rev.} {\bf D25}, 2649 (1982). 
%
\bibitem{FKV}
G.T. Fleming, J.B. Kogut, P.M. Vranas,
{\em Phys. Rev.} {\bf D64}, 034510 (2001).
%
\bibitem{CURVEN}
G. Curci and G. Veneziano,
{\em Nucl. Phys.} {\bf B292}, 555 (1987).
%
\bibitem{HMD}
S. Gottlieb, W. Liu, D. Toussaint, R.L. Renken, R.L. Sugar,
{\em Phys. Rev.} {\bf D35}, 2531 (1987).
%
\bibitem{TSMB}
I. Montvay,
{\em Nucl. Phys.} {\bf B466}, 259 (1996).
%
\bibitem{MB}
M. L\"uscher,
{\em Nucl. Phys.} {\bf B418}, 637 (1994).
%
\bibitem{OVERLAP}
H. Neuberger,
{\em Phys. Rev. Lett.} {\bf 81}, 4060 (1998).
%
\bibitem{BORFOR}
A. Borici, Ph. de Forcrand,
{\em Nucl. Phys.} {\bf B454}, 645 (1995). 
%
\bibitem{FREJAN}
R. Frezzotti and K. Jansen,
{\em Phys. Lett.} {\bf B402}, 328 (1997).
%
\bibitem{SPECTRUM}
I. Campos et al. (DESY-M\"unster Collaboration),
{\em Eur. Phys. J.} {\bf C11}, 507 (1999).
%
\bibitem{POLYNOM}
I. Montvay,
{\em Comput. Phys. Commun.} {\bf 109}, 144 (1998);  and
in {\em Numerical challenges in lattice quantum chromodynamics,
Wuppertal 1999}, Springer 2000, p. 153; hep-lat/9911014.
%
\bibitem{VARYACT}
I. Montvay,
hep-lat/0111015. 
%
\bibitem{PFAFFIAN}
N. Bourbaki,
{\em Alg\`ebre}, Chap.\ IX., Hermann, Paris, 1959.
%
\bibitem{BOULDER}
R. Kirchner et al. (DESY-M\"unster Collaboration),
{\em Nucl. Phys. Proc. Suppl.} {\bf 73}, 828 (1999).
%
\bibitem{SPECTRALFLOW}
R.G. Edwards, U.M. Heller, R. Narayanan,
{\em Nucl. Phys.} {\bf B535}, 403 (1998).
%
\bibitem{WTIB}
F. Farchioni et al. (DESY-M\"unster-Roma Collaboration),
{\em Nucl. Phys. Proc. Suppl.} {\bf 94}, 787 (2001).
%
\bibitem{WTI}
F. Farchioni et al. (DESY-M\"unster-Roma Collaboration),
hep-lat/0111008.
%
\bibitem{DESYSWANSEA}
S. Hands, I. Montvay, S. Morrison, M. Oevers, L. Scorzato, J. Skullerud,
{\em Eur. Phys. J.} {\bf C17}, 285 (2000).
%
\bibitem{HMC}
S. Duane, A.D. Kennedy, B.J. Pendleton, D. Roweth,
{\em Phys. Lett.} {\bf B195}, 216 (1987).
%
\bibitem{DISCHIRAL}
R. Kirchner et al. (DESY-M\"unster Collaboration),
{\em Phys. Lett.} {\bf B446}, 209 (1999).
%
\bibitem{SINT}
S. Sint,
{\em Nucl. Phys. Proc. Suppl.} {\bf 94}, 79 (2001).
%
\bibitem{TANIG}
Y. Taniguchi,
{\em Phys. Rev.} {\bf D63}, 014502 (2001).
%
\bibitem{SUSYWTP}
F. Farchioni et al. (DESY-M\"unster Collaboration),
{\em Nucl. Phys. Proc. Suppl.} {\bf 94}, 791 (2001);
hep-lat/0110113.
%
\bibitem{PISA}
A. Feo et al. (DESY-M\"unster Collaboration),
{\em Nucl. Phys. Proc. Suppl.} {\bf 83}, 670 (2000).
%
\bibitem{KIRCHNER}
R. Kirchner,
{\em Ward identities and mass spectrum of N=1 Super Yang-Mills
theory on the lattice,}
PhD Thesis, University Hamburg, 2000.
%
\bibitem{KUFUMIOKUK}
Y. Kuramashi, M. Fukugita, H. Mino, M. Okawa, A. Ukawa,
{\em Phys. Rev. Lett.} {\bf 72}, 3448 (1994).
%
\bibitem{BMMRT}
M. Bochicchio, L. Maiani, G. Martinelli, G. Rossi, M. Testa,
{\em Nucl. Phys.} {\bf B262}, 331 (1985).
%
\bibitem{SOMMER}
R. Sommer, {\em Nucl. Phys.} {\bf B411}, 839 (1994).
%
\end{thebibliography}
\end{document}